\documentclass{article}
\usepackage{amssymb,amsmath,amsfonts,graphicx,hyperref}
\newcommand{\mcm}[3]{\newcommand{#1}[#2]{{\ensuremath{#3}}}}

\usepackage{latexsym}
\usepackage{amssymb}

\mcm{\blank}{0}{(\emptybk)} \mcm{\dashbk}{0}{\mbox{---}}
\mcm{\emptybk}{0}{\:\:} \mcm{\hyph}{0}{\mbox{-}}

\mcm{\diagspace}{0}{\mbox{\hspace{2em}}}

\mcm{\cat}{1}{\mc{#1}} \mcm{\fcat}{1}{\mb{#1}}
\mcm{\mc}{1}{\mathcal{#1}} \mcm{\mr}{1}{\mathrm{#1}}
\mcm{\mi}{1}{\mathit{#1}} \mcm{\mb}{1}{\mathbf{#1}}
\mcm{\scat}{1}{\Bbb{#1}} \mcm{\twid}{1}{\widetilde{#1}}

\mcm{\elt}{0}{\in} \mcm{\sub}{0}{\,\subseteq\,}
\mcm{\such}{0}{\:|\:} \mcm{\without}{0}{\setminus}

\mcm{\atsr}{0}{\Box} \mcm{\eqv}{0}{\,\simeq\,}
\mcm{\iso}{0}{\,\cong\,}
\mcm{\of}{0}{\raisebox{0.2mm}{\ensuremath{\scriptstyle\circ}}}

\mcm{\bdry}{0}{\partial}

\mcm{\Bee}{0}{\cat{B}} \mcm{\Beep}{0}{\cat{B'}}
\mcm{\Eee}{0}{\cat{E}} \mcm{\Eeep}{0}{\cat{E'}}

\mcm{\Ess}{0}{\cat{S}} \mcm{\Tee}{0}{\cat{T}}
\mcm{\Teep}{0}{\cat{T'}} \mcm{\Stee}{0}{\scat{T}}
\mcm{\Steep}{0}{\scat{T'}}

\mcm{\blbk}{0}{\blank^{\blob}}
\mcm{\blob}{0}{\scriptscriptstyle{\bullet}}
\mcm{\stbk}{0}{\blank^{*}} \mcm{\ubl}{0}{{}^{\blob}}
\mcm{\ust}{0}{{}^{*}}

\mcm{\Cartpr}{0}{\pr{\Eee}{T}} \mcm{\Cartprp}{0}{\pr{\Eeep}{T'}}
\mcm{\Mnd}{0}{\triple{T}{\eta}{\mu}}
\mcm{\Zeropr}{0}{\pr{\Set}{\id}}

\mcm{\dopset}{0}{\ftrcat{\Delta^{\op}}{\Set}}
\mcm{\tropset}{0}{\ftrcat{\fcat{TR}^{\op}}{\Set}}

\mcm{\cod}{0}{\mr{cod}} \mcm{\dom}{0}{\mr{dom}}
\mcm{\End}{0}{\mr{End}} \mcm{\Hom}{0}{\mr{Hom}}
\mcm{\ob}{0}{\mr{ob}\,} \mcm{\op}{0}{\mr{op}}

\mcm{\comp}{0}{\mi{comp}} \mcm{\id}{0}{\mi{id}}
\mcm{\ids}{0}{\mi{ids}} \mcm{\mult}{0}{\mi{mult}}
\mcm{\unit}{0}{\mi{unit}}

\mcm{\Ab}{0}{\fcat{Ab}} \mcm{\Alg}{0}{\fcat{Alg}}
\mcm{\Bim}{1}{\fcat{Bim}(#1)} \mcm{\Cat}{0}{\fcat{Cat}}
\mcm{\Cay}{0}{\fcat{Cay}} \mcm{\Cpn}{1}{\pr{\Set/S_{#1}}{T_{#1}}}
\mcm{\fc}{0}{\fcat{fc}} \mcm{\fm}{0}{\fcat{fm}}
\mcm{\Graph}{0}{\fcat{Graph}} \mcm{\Gy}{0}{\fcat{Gy}}
\mcm{\Hpn}{1}{\pr{\Eee_{#1}}{P_{#1}}} \mcm{\Mon}{0}{\mb{Mon}}
\mcm{\Multicat}{0}{\fcat{Multicat}} \mcm{\One}{0}{\fcat{1}}
\mcm{\PD}{1}{\fcat{PD}_{#1}} \mcm{\Prof}{0}{\fcat{Prof}}
\mcm{\Set}{0}{\fcat{Set}} \mcm{\Span}{0}{\fcat{Span}}
\mcm{\Ssq}{0}{\fcat{Ssq}} \mcm{\Struc}{0}{\fcat{Struc}}
\mcm{\Sym}{0}{\fcat{Sym}} \mcm{\TR}{1}{\fcat{TR}(#1)}
\mcm{\Tr}{0}{\fcat{Tr}} \mcm{\Twocat}{0}{\fcat{2\hyph\Cat}}

\mcm{\integers}{0}{\mathbb{Z}}

\mcm{\range}{2}{#1,\,\ldots\,,#2}
\mcm{\bftuple}{2}{\tuplebts{\range{#1}{#2}}}
\mcm{\tuple}{3}{\tuplebts{\range{#1,#2}{#3}}}
\mcm{\rttuple}{1}{\tuplebts{\,\ldots\,,#1}}
\mcm{\abftuple}{2}{\atuplebts{\range{#1}{#2}}}
\mcm{\atuple}{3}{\atuplebts{\range{#1,#2}{#3}}}
\mcm{\arttuple}{1}{\atuplebts{\,\ldots\,,#1}}
\mcm{\sqbftuple}{2}{\obt\range{#1}{#2}\cbt}
\mcm{\pr}{2}{\tuplebts{#1,#2}}
\mcm{\triple}{3}{\tuplebts{#1,#2,#3}}

\mcm{\eend}{2}{#1[#2]} \mcm{\ehom}{3}{#1[#2,#3]}
\mcm{\ftrcat}{2}{[#1,#2]} \mcm{\homset}{3}{#1(#2,#3)}
\mcm{\multihom}{3}{#1(#2;#3)}
\mcm{\relhom}{5}{#1_{#2}(\range{#3}{#4};#5)}

\mcm{\go}{0}{\rTo} \mcm{\goby}{1}{\rTo^{#1}}
\mcm{\goesto}{0}{\,\longmapsto\,} \mcm{\goiso}{0}{\goby{\diso}}
\mcm{\monic}{0}{\rMonic} \mcm{\og}{0}{\lTo}
\mcm{\ogby}{1}{\lTo^{#1}}

\mcm{\gph}{2}{\spn{#1}{T #2}{#2}} \mcm{\graph}{4}{\spaan{#1}{T
#2}{#2}{#3}{#4}} \mcm{\oppair}{2}{\stackrel{\rTo^{#1}}{\lTo_{#2}}}
\mcm{\parpair}{2}{\stackrel{\rTo^{#1}}{\rTo_{#2}}}
\mcm{\spn}{3}{#2 \og #1 \go #3} \mcm{\spaan}{5}{#2 \ogby{#4} #1
\goby{#5} #3}

\mcm{\bktdvslob}{3}
    {\left(
    \begin{diagram}[height=1.5em]
    #1      \\
    \dTo>{\,#2} \\
    #3      \\
    \end{diagram}
    \right)}
\mcm{\slob}{3}{(#1 \goby{#2} #3)} \mcm{\vslob}{3}
    {\left.
    \begin{diagram}[height=1.5em]
    #1      \\
    \dTo>{\,#2} \\
    #3      \\
    \end{diagram}
    \right.}

\newenvironment{tree}
    {\begin{diagram}[height=1em,width=.75em,abut,noPS,tight]}
    {\end{diagram}}

\mcm{\enode}{0}{\circ}

\mcm{\nl}{1}{\stackrel{\textstyle #1}{\node}}
\mcm{\node}{0}{\bullet}

\mcm{\utree}{0}{\node}

\mcm{\diso}{0}{\sim}

\mcm{\vdiso}{0}{\wr}

\mcm{\nat}{0}{\mathbb{N}}

\mcm{\Onepr}{0}{\pr{\Graph}{\fc}}
\newlength{\nllwidth}
\newlength{\nllheight}
\newcommand{\stackbelow}[2]{%
\settowidth{\nllwidth}{\ensuremath{#1}\ensuremath{#2}}%
\settoheight{\nllheight}{\ensuremath{#2}}%
\addtolength{\nllheight}{.3ex}%
\mbox{%
\ensuremath{#1}%
\hspace{-.5\nllwidth}%
\raisebox{-1\nllheight}{\ensuremath{#2}}}}
\mcm{\nlal}{2}{\stackbelow{\nl{#1}}{#2}}
\mcm{\nll}{1}{\stackbelow{\node}{#1}} \mcm{\wun}{0}{\fcat{1}}
\mcm{\atuplebts}{1}{\langle #1 \rangle} \mcm{\tuplebts}{1}{(#1)}
\mcm{\bo}{0}{(} \mcm{\bc}{0}{)}
\mcm{\UBilax}{0}{\fcat{UBicat}_\mr{lax}}
\mcm{\UBiwk}{0}{\fcat{UBicat}_\mr{wk}}
\mcm{\UBistr}{0}{\fcat{UBicat}_\mr{str}}
\mcm{\Bilax}{0}{\fcat{Bicat}_\mr{lax}}
\mcm{\Biwk}{0}{\fcat{Bicat}_\mr{wk}}
\mcm{\Bistr}{0}{\fcat{Bicat}_\mr{str}} \mcm{\rotsub}{0}{\cup
\raisebox{0.1em}{$\scriptstyle{|}$}} \mcm{\pd}{0}{\fcat{pd}}
\mcm{\rep}{1}{\widehat{#1}} \mcm{\ovln}{1}{\overline{#1}}
\mcm{\Gph}{0}{\fcat{Gph}} \mcm{\tr}{0}{\fcat{tr}}

\mcm{\ladj}{0}{\,\dashv\,} \mcm{\zeropd}{0}{\node}
    {\end{diagram}}
\mcm{\END}{0}{\fcat{End}} \mcm{\HOM}{0}{\fcat{Hom}}

\newlength{\gwidth} 
\newlength{\gvert}  
\newlength{\gdrop}  
\newlength{\gbaredrop}  
\newlength{\goffset}    
\newlength{\gtemp}  

\newcommand{\present}[1]{%
\makebox[1\gwidth]{%
\rule[-1\gdrop]{0ex}{1\gvert}%
\raisebox{-1\gbaredrop}{#1}}}

\newcommand{\presentl}[1]{%
\makebox[1\gwidth][l]{%
\rule[-1\gdrop]{0ex}{1\gvert}%
\raisebox{-1\gbaredrop}{#1}}}

\newcommand{\presentr}[1]{%
\makebox[1\gwidth][r]{%
\rule[-1\gdrop]{0ex}{1\gvert}%
\raisebox{-1\gbaredrop}{#1}}}

\newcommand{\ginitdims}[2]{
\setlength{\unitlength}{1em}
\setlength{\goffset}{.25\unitlength}
\setlength{\gwidth}{#1\unitlength}
\setlength{\gvert}{#2\unitlength}
\setlength{\gdrop}{.5\gvert}
\addtolength{\gdrop}{-1\goffset}
\setlength{\gbaredrop}{1\gdrop}
\addtolength{\gvert}{.6\unitlength}
\addtolength{\gdrop}{.3\unitlength}}    

\newcommand{\cinitdims}[2]{
\setlength{\unitlength}{1em}
\setlength{\goffset}{.35\unitlength}
\setlength{\gwidth}{#1\unitlength}
\setlength{\gvert}{#2\unitlength}
\setlength{\gdrop}{.5\gvert}
\addtolength{\gdrop}{-1\goffset}
\setlength{\gbaredrop}{1\gdrop}
\addtolength{\gvert}{.6\unitlength}
\addtolength{\gdrop}{.3\unitlength}}    

\newcommand{\gsinitdims}[2]{
\setlength{\unitlength}{0.5em}
\setlength{\goffset}{.25\unitlength}
\setlength{\gwidth}{#1\unitlength}
\setlength{\gvert}{#2\unitlength}
\setlength{\gdrop}{.5\gvert}
\addtolength{\gdrop}{-1\goffset}
\setlength{\gbaredrop}{1\gdrop}
\addtolength{\gvert}{.6\unitlength}
\addtolength{\gdrop}{.3\unitlength}}    

\newcommand{\sidespic}[1]{%
\settowidth{\gtemp}{\ensuremath{#1}}%
\addtolength{\gwidth}{1\gtemp}}

\newcommand{\abovepic}[1]{%
\settoheight{\gtemp}{\ensuremath{#1}}%
\addtolength{\gvert}{1\gtemp}%
\settodepth{\gtemp}{\ensuremath{#1}}%
\addtolength{\gvert}{1\gtemp}}

\newcommand{\belowpic}[1]{%
\settoheight{\gtemp}{\ensuremath{#1}}%
\addtolength{\gvert}{1\gtemp}%
\addtolength{\gdrop}{1\gtemp}%
\settodepth{\gtemp}{\ensuremath{#1}}%
\addtolength{\gvert}{1\gtemp}%
\addtolength{\gdrop}{1\gtemp}}

\newcommand{\cell}[4]{\put(#1,#2){\makebox(0,0)[#3]{\ensuremath{#4}}}}
\mcm{\zmark}{0}{\scriptstyle{\bullet}}

\newcommand{\pregfst}[1]{%
\begin{picture}(0.5,0.2)(-0.5,-0.2)%
\cell{-0.1}{-0.2}{tr}{#1}%
\cell{0}{0}{c}{\zmark}%
\end{picture}}

\mcm{\gfst}{1}{%
\ginitdims{0.5}{0.4}%
\sidespic{#1}%
\belowpic{#1}%
\presentr{\pregfst{#1}}}

\newcommand{\preglst}[1]{%
\begin{picture}(0.5,0.2)(0,-0.2)%
\cell{0.1}{-0.2}{tl}{#1}%
\cell{0.05}{0}{c}{\zmark}%
\end{picture}}

\mcm{\glst}{1}{%
\ginitdims{.5}{.4}%
\sidespic{#1}%
\belowpic{#1}%
\presentl{\preglst{#1}}}

\newcommand{\preglft}[1]{%
\begin{picture}(0,0.2)(0,-0.2)%
\cell{-0.1}{-0.2}{tr}{#1}%
\cell{0.05}{0}{c}{\zmark}%
\end{picture}}

\mcm{\glft}{1}{%
\ginitdims{0}{.4}%
\belowpic{#1}%
\present{\preglft{#1}}}

\newcommand{\pregrgt}[1]{%
\begin{picture}(0,0.2)(0,-0.2)%
\cell{0.1}{-0.2}{tl}{#1}%
\cell{0.05}{0}{c}{\zmark}%
\end{picture}}

\mcm{\grgt}{1}{%
\ginitdims{0}{.4}%
\belowpic{#1}%
\present{\pregrgt{#1}}}

\newcommand{\pregblw}[1]{%
\begin{picture}(0,0.3)(0,-0.3)
\cell{0}{-0.3}{t}{#1}%
\cell{0.05}{0}{c}{\zmark}%
\end{picture}}

\mcm{\gblw}{1}{%
\ginitdims{0}{.6}%
\belowpic{#1}%
\present{\pregblw{#1}}}

\newcommand{\pregfbw}[1]{%
\begin{picture}(0,0.65)(0,-0.65)
\cell{0}{-0.65}{t}{#1}%
\cell{0.05}{0}{c}{\zmark}%
\end{picture}}

\mcm{\gfbw}{1}{%
\ginitdims{0}{1.3}%
\belowpic{#1}%
\present{\pregfbw{#1}}}

\newcommand{\pregzero}[1]{%
\begin{picture}(0.8,0.4)(-0.4,-0.4)
\cell{0}{-0.4}{t}{#1}%
\cell{0}{0}{c}{\zmark}%
\end{picture}}

\mcm{\gzero}{1}{%
\ginitdims{0.8}{.6}%
\belowpic{#1}%
\sidespic{#1}%
\present{\pregzero{#1}}}

\newcommand{\pregone}[1]{%
\begin{picture}(5,0.4)(0,-0.2)%
\cell{2.5}{0.2}{b}{#1}%
\put(0,0){\vector(1,0){5}}%
\end{picture}}

\mcm{\gone}{1}{%
\ginitdims{5}{0.4}%
\abovepic{#1}%
\present{\pregone{#1}}}

\newcommand{\pregtwo}[3]{%
\begin{picture}(5,3.4)(0,-0.2)%
\cell{2.5}{3.2}{b}{#1}%
\cell{2.5}{-.2}{t}{#2}%
\cell{2.7}{1.5}{l}{#3}%
\qbezier(0,1.5)(2.5,4.5)(5,1.5)%
\qbezier(0,1.5)(2.5,-1.5)(5,1.5)%
\put(5,1.5){\vector(1,-1){0}}%
\put(5,1.5){\vector(1,1){0}}%
\put(2.5,2.5){\vector(0,-1){2}}%
\end{picture}}

\mcm{\gtwo}{3}{%
\ginitdims{5}{3.4}%
\abovepic{#1}%
\belowpic{#2}%
\present{\pregtwo{#1}{#2}{#3}}}

\newcommand{\pregthree}[5]{%
\begin{picture}(5,5.4)(0,-1.2)%
\cell{2.5}{4.2}{b}{#1}%
\cell{1.5}{1.7}{b}{#2}%
\cell{2.5}{-1.2}{t}{#3}%
\cell{2.7}{2.75}{l}{#4}%
\cell{2.7}{0.25}{l}{#5}%
\qbezier(0,1.5)(2.5,6.5)(5,1.5)%
\qbezier(0,1.5)(2.5,-3.5)(5,1.5)%
\put(0,1.5){\vector(1,0){5}}%
\put(2.5,3.5){\vector(0,-1){1.5}}%
\put(2.5,1){\vector(0,-1){1.5}}%
\put(5,1.5){\vector(1,-3){0}}%
\put(5,1.5){\vector(1,3){0}}%
\end{picture}}

\mcm{\gthree}{5}{%
\ginitdims{5}{5.4}%
\abovepic{#1}%
\belowpic{#3}%
\present{\pregthree{#1}{#2}{#3}{#4}{#5}}}

\newcommand{\pregfour}[7]{%
\begin{picture}(5,8.4)(0,-2.7)%
\cell{2.5}{5.7}{b}{#1}%
\cell{1.5}{2.8}{b}{#2}%
\cell{1.5}{0.2}{t}{#3}%
\cell{2.5}{-2.7}{t}{#4}%
\cell{2.7}{4.25}{l}{#5}%
\cell{2.7}{1.5}{l}{#6}%
\cell{2.7}{-1.25}{l}{#7}%
\qbezier(0,1.5)(2.5,9.5)(5,1.5)%
\qbezier(0,1.5)(2.5,4)(5,1.5)%
\qbezier(0,1.5)(2.5,-1)(5,1.5)%
\qbezier(0,1.5)(2.5,-6.5)(5,1.5)%
\put(2.5,5.25){\vector(0,-1){2}}%
\put(2.5,2.5){\vector(0,-1){2}}%
\put(2.5,-0.25){\vector(0,-1){2}}%
\put(5,1.5){\vector(1,-4){0}}%
\put(5,1.5){\vector(4,-3){0}}%
\put(5,1.5){\vector(4,3){0}}%
\put(5,1.5){\vector(1,4){0}}%
\end{picture}}

\mcm{\gfour}{7}{%
\ginitdims{5}{8.4}%
\abovepic{#1}%
\belowpic{#4}%
\present{\pregfour{#1}{#2}{#3}{#4}{#5}{#6}{#7}}}

\newcommand{\pregthreecell}[5]{%
\begin{picture}(8,5)(-4,-2.5)%
\cell{0}{2.5}{b}{#1}%
\cell{0}{-2.5}{t}{#2}%
\cell{-1.7}{0}{r}{#3}%
\cell{1.7}{0}{l}{#4}%
\cell{0}{0.2}{b}{#5}%
\qbezier(-4,0)(0,4.2)(4,0)%
\qbezier(-4,0)(0,-4.2)(4,0)%
\qbezier(-0.5,1.8)(-2.5,0)(-0.5,-1.8)%
\qbezier(0.5,1.8)(2.5,0)(0.5,-1.8)%
\put(-1,0){\vector(1,0){2}}%
\put(4,0){\vector(1,-1){0}}%
\put(4,0){\vector(1,1){0}}%
\put(-0.5,-1.8){\vector(1,-1){0}}%
\put(0.5,-1.8){\vector(-1,-1){0}}%
\end{picture}}

\mcm{\gthreecell}{5}{%
\ginitdims{8}{5}%
\abovepic{#1}%
\belowpic{#2}%
\present{\pregthreecell{#1}{#2}{#3}{#4}{#5}}}

\newcommand{\pregthreecellu}{%
\begin{picture}(5,3.4)(-0.5,-0.2)%
\qbezier(-.5,1.5)(2,4.5)(4.5,1.5)%
\qbezier(-.5,1.5)(2,-1.5)(4.5,1.5)%
\qbezier(1.5,2.7)(0.5,1.5)(1.5,0.3)%
\qbezier(2.5,2.7)(3.5,1.5)(2.5,0.3)%
\put(1.3,1.5){\vector(1,0){1.4}}%
\put(4.5,1.5){\vector(1,-1){0}}%
\put(4.5,1.5){\vector(1,1){0}}%
\put(1.5,0.3){\vector(2,-3){0}}%
\put(2.5,0.3){\vector(-2,-3){0}}%
\end{picture}}

\mcm{\gthreecellu}{0}{%
\ginitdims{5}{3.4}%
\present{\pregthreecellu}}

\newcommand{\pregtwocentre}[3]{%
\begin{picture}(5,3.4)(0,-0.2)%
\cell{2.5}{3.2}{b}{#1}%
\cell{2.5}{-.2}{t}{#2}%
\cell{2.5}{1.5}{c}{#3}%
\qbezier(0,1.5)(2.5,4.5)(5,1.5)%
\qbezier(0,1.5)(2.5,-1.5)(5,1.5)%
\put(5,1.5){\vector(1,-1){0}}%
\put(5,1.5){\vector(1,1){0}}%
\put(2.5,2.5){\vector(0,-1){2}}%
\end{picture}}

\mcm{\gtwocentre}{3}{%
\ginitdims{5}{3.4}%
\abovepic{#1}%
\belowpic{#2}%
\present{\pregtwocentre{#1}{#2}{#3}}}

\newcommand{\pregspecialone}[9]{%
\begin{picture}(8,8)(-4,-4)%
\cell{0}{3.9}{b}{#1}%
\cell{-2}{-0.2}{t}{#2}%
\cell{0}{-3.9}{t}{#3}%
\cell{-1.5}{1.1}{r}{#4}%
\cell{0.2}{1.5}{l}{#5}%
\cell{1.5}{1.1}{l}{#6}%
\cell{0.2}{-2}{l}{#7}%
\cell{-0.9}{2.3}{b}{#8}%
\cell{0.9}{2.3}{b}{#9}%
\qbezier(-4,0)(0,8)(4,0)%
\qbezier(-4,0)(0,-8)(4,0)%
\qbezier(-0.5,3.4)(-3.5,2)(-0.5,0.6)%
\qbezier(0.5,3.4)(3.5,2)(0.5,0.6)%
\put(-4,0){\vector(1,0){8}}%
\put(0,3.4){\vector(0,-1){2.8}}%
\put(0,-0.8){\vector(0,-1){2.4}}%
\put(-1.5,2.2){\vector(1,0){1.2}}%
\put(0.3,2.2){\vector(1,0){1.2}}%
\put(4,0){\vector(1,-2){0}}%
\put(4,0){\vector(1,2){0}}%
\put(-0.5,0.6){\vector(2,-1){0}}%
\put(0.5,0.6){\vector(-2,-1){0}}%
\end{picture}}

\mcm{\gspecialone}{9}{%
\ginitdims{8}{8}%
\abovepic{#1}%
\belowpic{#3}%
\present{\pregspecialone{#1}{#2}{#3}{#4}{#5}{#6}{#7}{#8}{#9}}}

\newcommand{\pregspecialtwo}{%
\begin{picture}(5,3.4)(0,-0.2)%
\qbezier(0,1.5)(2.5,4.5)(5,1.5)%
\qbezier(0,1.5)(2.5,-1.5)(5,1.5)%
\qbezier(1.7,2.5)(0,1.5)(1.7,0.5)%
\qbezier(3.3,2.5)(5,1.5)(3.3,0.5)%
\put(5,1.5){\vector(1,-1){0}}%
\put(5,1.5){\vector(1,1){0}}%
\put(1.7,0.5){\vector(3,-2){0}}%
\put(3.3,0.5){\vector(-3,-2){0}}%
\put(2.5,2.5){\vector(0,-1){2}}%
\put(1.2,1.5){\vector(1,0){1}}%
\put(2.8,1.5){\vector(1,0){1}}%
\end{picture}}

\mcm{\gspecialtwo}{0}{%
\ginitdims{5}{3.4}%
\present{\pregspecialtwo}}

\newcommand{\pregspecialthree}{%
\begin{picture}(5,5.4)(0,-1.2)%
\qbezier(0,1.5)(2.5,6.5)(5,1.5)%
\qbezier(0,1.5)(2.5,-3.5)(5,1.5)%
\qbezier(2,3.5)(1,2.75)(2,2)%
\qbezier(3,3.5)(4,2.75)(3,2)%
\qbezier(2,1)(1,0.25)(2,-0.5)%
\qbezier(3,1)(4,0.25)(3,-0.5)%
\put(0,1.5){\vector(1,0){5}}%
\put(1.5,2.75){\vector(1,0){2}}%
\put(1.5,0.25){\vector(1,0){2}}%
\put(5,1.5){\vector(1,-3){0}}%
\put(5,1.5){\vector(1,3){0}}%
\put(2,2){\vector(1,-1){0}}%
\put(3,2){\vector(-1,-1){0}}%
\put(2,-0.5){\vector(1,-1){0}}%
\put(3,-0.5){\vector(-1,-1){0}}%
\end{picture}}

\mcm{\gspecialthree}{0}{%
\ginitdims{5}{5.4}%
\present{\pregspecialthree}}

\newcommand{\pregonew}[1]{%
\begin{picture}(8,0.4)(0,-0.2)%
\cell{4}{0.2}{b}{#1}%
\put(0,0){\vector(1,0){8}}%
\end{picture}}

\mcm{\gonew}{1}{%
\ginitdims{8}{0.4}%
\abovepic{#1}%
\present{\pregonew{#1}}}

\mcm{\gzersu}{0}{%
\gsinitdims{0}{.6}%
\present{\pregblw{}}}

\mcm{\gonesu}{0}{%
\gsinitdims{5}{0.4}%
\present{\pregone{}}}

\mcm{\gtwosu}{0}{%
\gsinitdims{5}{3.4}%
\present{\pregtwo{}{}{}}}

\mcm{\gthreesu}{0}{%
\gsinitdims{5}{5.4}%
\present{\pregthree{}{}{}{}{}}}

\mcm{\gfoursu}{0}{%
\gsinitdims{5}{8.4}%
\present{\pregfour{}{}{}{}{}{}{}}}

\newcommand{\precone}[1]{%
\begin{picture}(4.2,0.4)(-0.3,-0.2)%
\cell{1.8}{0.2}{b}{#1}%
\put(0,0){\vector(1,0){3.6}}%
\end{picture}}

\mcm{\cone}{1}{%
\cinitdims{4.2}{0.4}%
\abovepic{#1}%
\present{\precone{#1}}}

\mcm{\gfstsu}{0}{%
\gsinitdims{0.5}{0.4}%
\presentr{\pregfst{}}}

\mcm{\glstsu}{0}{%
\gsinitdims{0.5}{0.4}%
\presentl{\preglst{}}}

\newcommand{\prectwodbl}[3]%
{\begin{picture}(4.2,3.4)(-0.1,-0.2)%
\cell{2}{3.2}{b}{#1}%
\cell{2}{-0.2}{t}{#2}%
\cell{2.3}{1.5}{l}{#3}%
\qbezier(0,2)(2,4)(4,2)%
\qbezier(0,1)(2,-1)(4,1)%
\put(4,2){\vector(1,-1){0}}%
\put(4,1){\vector(1,1){0}}%
\put(1.9,2.5){\line(0,-1){1.8}}%
\put(2.1,2.5){\line(0,-1){1.8}}%
\cell{2.01}{0.4}{b}{\vee}%
\end{picture}}

\mcm{\ctwodbl}{3}{%
\cinitdims{4.2}{3.4}%
\abovepic{#1}%
\belowpic{#2}%
\present{\prectwodbl{#1}{#2}{#3}}}

\newcommand{\precthreedbl}[5]{%
\begin{picture}(4.2,5.4)(-0.1,-0.2)%
\cell{2}{5.2}{b}{#1}%
\cell{1}{2.7}{b}{#2}%
\cell{2}{-.2}{t}{#3}%
\cell{2.3}{3.75}{l}{#4}%
\cell{2.3}{1.25}{l}{#5}%
\qbezier(0,3)(2,7)(4,3)%
\qbezier(0,2)(2,-2)(4,2)%
\put(0,2.5){\vector(1,0){4}}%
\put(1.9,4.5){\line(0,-1){1.3}}%
\put(2.1,4.5){\line(0,-1){1.3}}%
\cell{2.01}{2.9}{b}{\vee}%
\put(1.9,2){\line(0,-1){1.3}}%
\put(2.1,2){\line(0,-1){1.3}}%
\cell{2.01}{0.4}{b}{\vee}%
\put(4,3){\vector(1,-3){0}}%
\put(4,2){\vector(1,3){0}}%
\end{picture}}

\mcm{\cthreedbl}{5}{%
\cinitdims{4.2}{5.4}%
\abovepic{#1}%
\belowpic{#3}%
\present{\precthreedbl{#1}{#2}{#3}{#4}{#5}}}

\newcommand{\precthreecelltrp}[5]{%
\begin{picture}(8.2,5)(-4.1,-2.5)%
\cell{0}{2.5}{b}{#1}%
\cell{0}{-2.5}{t}{#2}%
\cell{-1.8}{0}{r}{#3}%
\cell{1.8}{0}{l}{#4}%
\cell{0}{0.3}{b}{#5}%
\qbezier(-4,0.5)(0,4)(4,0.5)%
\qbezier(-4,-0.5)(0,-4)(4,-0.5)%
\qbezier(-0.6,2)(-2.6,0)(-0.6,-2)%
\qbezier(-0.4,2)(-2.4,0)(-0.5,-1.9)%
\cell{-0.6}{-2}{b}{\lrcorner}%
\qbezier(0.4,2)(2.4,0)(0.5,-1.9)%
\qbezier(0.6,2)(2.6,0)(0.6,-2)%
\cell{0.65}{-2}{b}{\llcorner}%
\put(-1,0.15){\line(1,0){1.7}}%
\put(-1,0){\line(1,0){2}}%
\put(-1,-0.15){\line(1,0){1.7}}%
\cell{1.15}{0}{r}{>}%
\put(4,0.5){\vector(1,-1){0}}%
\put(4,-0.5){\vector(1,1){0}}%
\end{picture}}

\mcm{\cthreecelltrp}{5}{%
\cinitdims{8.2}{5}%
\abovepic{#1}%
\belowpic{#2}%
\present{\precthreecelltrp{#1}{#2}{#3}{#4}{#5}}}

\newcommand{\prectwo}[3]%
{\begin{picture}(4.2,3.4)(-0.1,-0.2)%
\cell{2}{3.2}{b}{#1}%
\cell{2}{-0.2}{t}{#2}%
\cell{2.2}{1.5}{l}{#3}%
\qbezier(0,2)(2,4)(4,2)%
\qbezier(0,1)(2,-1)(4,1)%
\put(4,2){\vector(1,-1){0}}%
\put(4,1){\vector(1,1){0}}%
\put(2,2.5){\vector(0,-1){2}}%
\end{picture}}

\mcm{\ctwo}{3}{%
\cinitdims{4.2}{3.4}%
\abovepic{#1}%
\belowpic{#2}%
\present{\prectwo{#1}{#2}{#3}}}

\newcommand{\precthree}[5]{%
\begin{picture}(4.2,5.4)(-0.1,-0.2)%
\cell{2}{5.2}{b}{#1}%
\cell{1}{2.7}{b}{#2}%
\cell{2}{-.2}{t}{#3}%
\cell{2.2}{3.75}{l}{#4}%
\cell{2.2}{1.25}{l}{#5}%
\qbezier(0,3)(2,7)(4,3)%
\qbezier(0,2)(2,-2)(4,2)%
\put(0,2.5){\vector(1,0){4}}%
\put(2,4.5){\vector(0,-1){1.5}}%
\put(2,2){\vector(0,-1){1.5}}%
\put(4,3){\vector(1,-3){0}}%
\put(4,2){\vector(1,3){0}}%
\end{picture}}

\mcm{\cthree}{5}{%
\cinitdims{4.2}{5.4}%
\abovepic{#1}%
\belowpic{#3}%
\present{\precthree{#1}{#2}{#3}{#4}{#5}}}

\newcommand{\prectwoop}[3]%
{\begin{picture}(4.2,3.4)(-0.1,-0.2)%
\cell{2}{3.2}{b}{#1}%
\cell{2}{-0.2}{t}{#2}%
\cell{2.2}{1.5}{l}{#3}%
\qbezier(0,2)(2,4)(4,2)%
\qbezier(0,1)(2,-1)(4,1)%
\put(0,2){\vector(-1,-1){0}}%
\put(0,1){\vector(-1,1){0}}%
\put(2,2.5){\vector(0,-1){2}}%
\end{picture}}

\mcm{\ctwoop}{3}{%
\cinitdims{4.2}{3.4}%
\abovepic{#1}%
\belowpic{#2}%
\present{\prectwoop{#1}{#2}{#3}}}

\newcommand{\prectwopar}[4]{%
\begin{picture}(4.2,3.4)(-0.1,-0.2)%
\cell{2}{3.2}{b}{#1}%
\cell{2}{-0.2}{t}{#2}%
\cell{1.6}{1.5}{r}{#3}%
\cell{2.4}{1.5}{l}{#4}%
\qbezier(0,2)(2,4)(4,2)%
\qbezier(0,1)(2,-1)(4,1)%
\put(4,2){\vector(1,-1){0}}%
\put(4,1){\vector(1,1){0}}%
\put(1.8,2.5){\vector(0,-1){2}}%
\put(2.2,2.5){\vector(0,-1){2}}%
\end{picture}}

\mcm{\ctwopar}{4}{%
\cinitdims{4.2}{3.4}%
\abovepic{#1}%
\belowpic{#2}%
\present{\prectwopar{#1}{#2}{#3}{#4}}}

\newcommand{\precthreein}[5]{%
\begin{picture}(4.2,5.4)(-0.1,-0.2)%
\cell{2}{5.2}{b}{#1}%
\cell{1}{2.7}{b}{#2}%
\cell{2}{-.2}{t}{#3}%
\cell{2.2}{3.75}{l}{#4}%
\cell{2.2}{1.25}{l}{#5}%
\qbezier(0,3)(2,7)(4,3)%
\qbezier(0,2)(2,-2)(4,2)%
\put(0,2.5){\vector(1,0){4}}%
\put(2,4.5){\vector(0,-1){1.5}}%
\put(2,0.5){\vector(0,1){1.5}}%
\put(4,3){\vector(1,-3){0}}%
\put(4,2){\vector(1,3){0}}%
\end{picture}}

\mcm{\cthreein}{5}{%
\cinitdims{4.2}{5.4}%
\abovepic{#1}%
\belowpic{#3}%
\present{\precthreein{#1}{#2}{#3}{#4}{#5}}}

\newcommand{\precthreecell}[5]{%
\begin{picture}(8.2,5)(-4.1,-2.5)%
\cell{0}{2.5}{b}{#1}%
\cell{0}{-2.5}{t}{#2}%
\cell{-1.7}{0}{r}{#3}%
\cell{1.7}{0}{l}{#4}%
\cell{0}{0.2}{b}{#5}%
\qbezier(-4,0.5)(0,4)(4,0.5)%
\qbezier(-4,-0.5)(0,-4)(4,-0.5)%
\qbezier(-0.5,2)(-2.5,0)(-0.5,-2)%
\qbezier(0.5,2)(2.5,0)(0.5,-2)%
\put(-1,0){\vector(1,0){2}}%
\put(4,0.5){\vector(1,-1){0}}%
\put(4,-0.5){\vector(1,1){0}}%
\put(-0.5,-2){\vector(1,-1){0}}%
\put(0.5,-2){\vector(-1,-1){0}}%
\end{picture}}

\mcm{\cthreecell}{5}{%
\cinitdims{8.2}{5}%
\abovepic{#1}%
\belowpic{#2}%
\present{\precthreecell{#1}{#2}{#3}{#4}{#5}}}

\newcommand{\precthreecellpar}[6]{%
\begin{picture}(8.2,5)(-4.1,-2.5)%
\cell{0}{2.5}{b}{#1}%
\cell{0}{-2.5}{t}{#2}%
\cell{-1.7}{0}{r}{#3}%
\cell{1.7}{0}{l}{#4}%
\cell{0}{0.4}{b}{#5}%
\cell{0}{-0.4}{t}{#6}%
\qbezier(-4,0.5)(0,4)(4,0.5)%
\qbezier(-4,-0.5)(0,-4)(4,-0.5)%
\qbezier(-0.5,2)(-2.5,0)(-0.5,-2)%
\qbezier(0.5,2)(2.5,0)(0.5,-2)%
\put(-1,0.2){\vector(1,0){2}}%
\put(-1,-0.2){\vector(1,0){2}}%
\put(4,0.5){\vector(1,-1){0}}%
\put(4,-0.5){\vector(1,1){0}}%
\put(-0.5,-2){\vector(1,-1){0}}%
\put(0.5,-2){\vector(-1,-1){0}}%
\end{picture}}

\mcm{\cthreecellpar}{6}{%
\cinitdims{8.2}{5}%
\abovepic{#1}%
\belowpic{#2}%
\present{\precthreecellpar{#1}{#2}{#3}{#4}{#5}{#6}}}

\newcommand{\prectwov}[5]{%
\begin{picture}(3.4,4.2)(0.8,0.9)%
\cell{2.5}{5.1}{b}{#1}%
\cell{2.5}{0.9}{t}{#2}%
\cell{0.8}{3}{r}{#3}%
\cell{4.2}{3}{l}{#4}%
\cell{2.5}{3.2}{b}{#5}%
\qbezier(2,5)(0,3)(2,1)%
\qbezier(3,5)(5,3)(3,1)%
\put(2,1){\vector(1,-1){0}}%
\put(3,1){\vector(-1,-1){0}}%
\put(1.5,3){\vector(1,0){2}}%
\end{picture}}

\mcm{\ctwov}{5}{%
\cinitdims{3.4}{4.2}%
\abovepic{#1}%
\belowpic{#2}%
\sidespic{#3}%
\sidespic{#4}%
\present{\prectwov{#1}{#2}{#3}{#4}{#5}}}

\newcommand{\precthreecellv}[7]{%
\begin{picture}(5,8.2)(0.5,-1.6)%
\cell{3}{6.6}{b}{#1}%
\cell{3}{-1.6}{t}{#2}%
\cell{0.5}{2.5}{r}{#3}%
\cell{5.5}{2.5}{l}{#4}%
\cell{3}{4.2}{b}{#5}%
\cell{3}{0.8}{t}{#6}%
\cell{3.2}{2.5}{l}{#7}%
\qbezier(3.5,6.5)(7,2.5)(3.5,-1.5)%
\qbezier(2.5,6.5)(-1,2.5)(2.5,-1.5)%
\put(2.5,-1.5){\vector(1,-1){0}}%
\put(3.5,-1.5){\vector(-1,-1){0}}%
\qbezier(1,3)(3,5)(5,3)%
\qbezier(1,2)(3,0)(5,2)%
\put(5,3){\vector(1,-1){0}}%
\put(5,2){\vector(1,1){0}}%
\put(3,3.5){\vector(0,-1){2}}%
\end{picture}}

\mcm{\cthreecellv}{7}{%
\cinitdims{5}{8.2}%
\abovepic{#1}%
\belowpic{#2}%
\sidespic{#3}%
\sidespic{#4}%
\present{\precthreecellv{#1}{#2}{#3}{#4}{#5}{#6}{#7}}}

\newcommand{\pretopez}[2]{%
\begin{picture}(2.6,2.3)(-1.3,-2.2)%
\cell{0}{-2.2}{t}{#1}%
\cell{0}{-1.2}{c}{#2}%
\qbezier(0,0)(-2,-2)(0,-2)%
\qbezier(0,0)(2,-2)(0,-2)%
\put(0,0){\vector(-1,1){0}}%
\end{picture}}

\mcm{\topez}{2}{%
\ginitdims{2.6}{2.3}%
\belowpic{#1}%
\present{\pretopez{#1}{#2}}}

\newcommand{\pretopea}[3]{%
\begin{picture}(4,1.9)(-2,-0,2)%
\cell{0}{1.7}{b}{#1}%
\cell{0}{-0.2}{t}{#2}%
\cell{0}{0.7}{c}{#3}%
\qbezier(-2,0)(0,3)(2,0)%
\put(-2,0){\vector(1,0){4}}%
\put(2,0){\vector(2,-3){0}}%
\end{picture}}

\mcm{\topea}{3}{%
\ginitdims{4}{1.9}%
\abovepic{#1}%
\belowpic{#2}%
\present{\pretopea{#1}{#2}{#3}}}

\newcommand{\pretopeb}[4]{%
\begin{picture}(4,2.2)(-2,-0.2)%
\cell{-1.1}{1}{br}{#1}%
\cell{1.1}{1}{bl}{#2}%
\cell{0}{-0.2}{t}{#3}%
\cell{0}{0.8}{c}{#4}%
\put(-2,0){\vector(1,1){2}}%
\put(0,2){\vector(1,-1){2}}%
\put(-2,0){\vector(1,0){4}}%
\end{picture}}

\mcm{\topeb}{4}{%
\ginitdims{4}{2.2}%
\belowpic{#3}%
\present{\pretopeb{#1}{#2}{#3}{#4}}}

\newcommand{\pretopec}[5]{%
\begin{picture}(4,2.2)(-2,-0.2)%
\cell{-1.8}{1}{br}{#1}%
\cell{0}{2.2}{b}{#2}%
\cell{1.8}{1}{bl}{#3}%
\cell{0}{-0.2}{t}{#4}%
\cell{0}{0.8}{c}{#5}%
\put(-2,0){\vector(1,2){1}}%
\put(-1,2){\vector(1,0){2}}%
\put(1,2){\vector(1,-2){1}}%
\put(-2,0){\vector(1,0){4}}%
\end{picture}}

\mcm{\topec}{5}{%
\ginitdims{4}{2.2}%
\sidespic{#1}%
\abovepic{#2}%
\sidespic{#3}%
\belowpic{#4}%
\present{\pretopec{#1}{#2}{#3}{#4}{#5}}}

\newcommand{\pretoped}[6]{%
\begin{picture}(4,2.5)(-2,-0.2)%
\cell{-2}{0.6}{br}{#1}%
\cell{-0.7}{2.2}{br}{#2}%
\cell{0.7}{2.2}{bl}{#3}%
\cell{2}{0.6}{bl}{#4}%
\cell{0}{-0.2}{t}{#5}%
\cell{0}{0.8}{c}{#6}%
\put(-2,0){\vector(1,3){0.5}}%
\put(-1.5,1.5){\vector(3,2){1.5}}%
\put(0,2.5){\vector(3,-2){1.5}}%
\put(1.5,1.5){\vector(1,-3){0.5}}%
\put(-2,0){\vector(1,0){4}}%
\end{picture}}

\mcm{\toped}{6}{%
\ginitdims{4}{2.5}%
\sidespic{#1}%
\abovepic{#2}%
\abovepic{#3}%
\sidespic{#4}%
\belowpic{#5}%
\present{\pretoped{#1}{#2}{#3}{#4}{#5}{#6}}}

\newcommand{\pretopeq}[5]{%
\begin{picture}(4,2.5)(-2,-0.2)%
\cell{-2}{0.6}{br}{#1}%
\cell{-1}{2.2}{br}{#2}%
\cell{2}{0.6}{bl}{#3}%
\cell{0}{-0.2}{t}{#4}%
\cell{0}{0.8}{c}{#5}%
\put(-2,0){\vector(1,3){0.5}}%
\put(-1.5,1.5){\vector(1,1){1}}%
\cell{0.9}{2.3}{c}{\ddots}
\put(1.5,1.5){\vector(1,-3){0.5}}%
\put(-2,0){\vector(1,0){4}}%
\end{picture}}

\mcm{\topeq}{5}{%
\ginitdims{4}{2.5}%
\sidespic{#1}%
\abovepic{#2}%
\sidespic{#3}%
\belowpic{#4}%
\present{\pretopeq{#1}{#2}{#3}{#4}{#5}}}

\newcommand{\pretopebase}[1]{%
\begin{picture}(4,0.4)(0,-0.2)%
\cell{2}{0.2}{b}{#1}%
\put(0,0){\vector(1,0){4}}%
\end{picture}}

\mcm{\topebase}{1}{%
\ginitdims{4}{0.4}%
\abovepic{#1}%
\present{\pretopebase{#1}}}

\newcommand{\pretopezs}[2]{%
\begin{picture}(2.6,2.3)(-1.3,-2.2)%
\cell{0}{-2.2}{t}{#1}%
\cell{0}{-1.2}{c}{#2}%
\qbezier(0,0)(-2,-2)(0,-2)%
\qbezier(0,0)(2,-2)(0,-2)%
\end{picture}}

\mcm{\topezs}{2}{%
\ginitdims{2.6}{2.3}%
\belowpic{#1}%
\present{\pretopezs{#1}{#2}}}

\newcommand{\pretopeas}[3]{%
\begin{picture}(4,1.9)(-2,-0,2)%
\cell{0}{1.7}{b}{#1}%
\cell{0}{-0.2}{t}{#2}%
\cell{0}{0.7}{c}{#3}%
\qbezier(-2,0)(0,3)(2,0)%
\put(-2,0){\line(1,0){4}}%
\end{picture}}

\mcm{\topeas}{3}{%
\ginitdims{4}{1.9}%
\abovepic{#1}%
\belowpic{#2}%
\present{\pretopeas{#1}{#2}{#3}}}

\newcommand{\pretopebs}[4]{%
\begin{picture}(4,2.2)(-2,-0.2)%
\cell{-1.1}{1}{br}{#1}%
\cell{1.1}{1}{bl}{#2}%
\cell{0}{-0.2}{t}{#3}%
\cell{0}{0.8}{c}{#4}%
\put(-2,0){\line(1,1){2}}%
\put(0,2){\line(1,-1){2}}%
\put(-2,0){\line(1,0){4}}%
\end{picture}}

\mcm{\topebs}{4}{%
\ginitdims{4}{2.2}%
\belowpic{#3}%
\present{\pretopebs{#1}{#2}{#3}{#4}}}

\newcommand{\pretopecs}[5]{%
\begin{picture}(4,2.2)(-2,-0.2)%
\cell{-1.8}{1}{br}{#1}%
\cell{0}{2.2}{b}{#2}%
\cell{1.8}{1}{bl}{#3}%
\cell{0}{-0.2}{t}{#4}%
\cell{0}{0.8}{c}{#5}%
\put(-2,0){\line(1,2){1}}%
\put(-1,2){\line(1,0){2}}%
\put(1,2){\line(1,-2){1}}%
\put(-2,0){\line(1,0){4}}%
\end{picture}}

\mcm{\topecs}{5}{%
\ginitdims{4}{2.2}%
\sidespic{#1}%
\abovepic{#2}%
\sidespic{#3}%
\belowpic{#4}%
\present{\pretopecs{#1}{#2}{#3}{#4}{#5}}}

\newcommand{\pretopeds}[6]{%
\begin{picture}(4,2.5)(-2,-0.2)%
\cell{-2}{0.6}{br}{#1}%
\cell{-0.7}{2.2}{br}{#2}%
\cell{0.7}{2.2}{bl}{#3}%
\cell{2}{0.6}{bl}{#4}%
\cell{0}{-0.2}{t}{#5}%
\cell{0}{0.8}{c}{#6}%
\put(-2,0){\line(1,3){0.5}}%
\put(-1.5,1.5){\line(3,2){1.5}}%
\put(0,2.5){\line(3,-2){1.5}}%
\put(1.5,1.5){\line(1,-3){0.5}}%
\put(-2,0){\line(1,0){4}}%
\end{picture}}

\mcm{\topeds}{6}{%
\ginitdims{4}{2.5}%
\sidespic{#1}%
\abovepic{#2}%
\abovepic{#3}%
\sidespic{#4}%
\belowpic{#5}%
\present{\pretopeds{#1}{#2}{#3}{#4}{#5}{#6}}}

\newcommand{\pretopeqs}[5]{%
\begin{picture}(4,2.5)(-2,-0.2)%
\cell{-2}{0.6}{br}{#1}%
\cell{-1}{2.2}{br}{#2}%
\cell{2}{0.6}{bl}{#3}%
\cell{0}{-0.2}{t}{#4}%
\cell{0}{0.8}{c}{#5}%
\put(-2,0){\line(1,3){0.5}}%
\put(-1.5,1.5){\line(1,1){1}}%
\cell{0.9}{2.3}{c}{\ddots}
\put(1.5,1.5){\line(1,-3){0.5}}%
\put(-2,0){\line(1,0){4}}%
\end{picture}}

\mcm{\topeqs}{5}{%
\ginitdims{4}{2.5}%
\sidespic{#1}%
\abovepic{#2}%
\sidespic{#3}%
\belowpic{#4}%
\present{\pretopeqs{#1}{#2}{#3}{#4}{#5}}}

\newcommand{\pretopebases}[1]{%
\begin{picture}(4,0.4)(0,-0.2)%
\cell{2}{0.2}{b}{#1}%
\put(0,0){\line(1,0){4}}%
\end{picture}}

\mcm{\topebases}{1}{%
\ginitdims{4}{0.4}%
\abovepic{#1}%
\present{\pretopebases{#1}}}

\newcommand{\pregdots}[6]{%
\begin{picture}(5,8.4)(0,-2.7)%
\cell{2.5}{5.7}{b}{#1}%
\cell{1.5}{2.8}{b}{#2}%
\cell{1.5}{0.2}{t}{#3}%
\cell{2.5}{-2.7}{t}{#4}%
\cell{2.7}{4.25}{l}{#5}%
\cell{2.7}{-1.25}{l}{#6}%
\qbezier(0,1.5)(2.5,9.5)(5,1.5)%
\qbezier(0,1.5)(2.5,4)(5,1.5)%
\qbezier(0,1.5)(2.5,-1)(5,1.5)%
\qbezier(0,1.5)(2.5,-6.5)(5,1.5)%
\put(2.5,5.25){\vector(0,-1){2}}%
\put(2.5,-0.25){\vector(0,-1){2}}%
\cell{2.5}{1.7}{c}{\vdots}%
\put(5,1.5){\vector(1,-4){0}}%
\put(5,1.5){\vector(4,-3){0}}%
\put(5,1.5){\vector(4,3){0}}%
\put(5,1.5){\vector(1,4){0}}%
\end{picture}}

\mcm{\gdots}{6}{%
\ginitdims{5}{8.4}%
\abovepic{#1}%
\belowpic{#4}%
\present{\pregdots{#1}{#2}{#3}{#4}{#5}{#6}}}

\newlength{\volt}
\makeatletter

\def\diagram{\m@th\leftwidth=\z@ \rightwidth=\z@ \topheight=\z@
\botheight=\z@ \setbox\@picbox\hbox\bgroup}

\def\enddiagram{\egroup\wd\@picbox\rightwidth\unitlength
\ht\@picbox\topheight\unitlength \dp\@picbox\botheight\unitlength
\hskip\leftwidth\unitlength\box\@picbox}

\def\bfig{\begin{diagram}}
\def\efig{\end{diagram}}
\newcount\wideness \newcount\leftwidth \newcount\rightwidth
\newcount\highness \newcount\topheight \newcount\botheight

\def\ratchet#1#2{\ifnum#1<#2 \global #1=#2 \fi}

\def\putbox(#1,#2)#3{%
\horsize{\wideness}{#3} \divide\wideness by 2 {\advance\wideness
by #1 \ratchet{\rightwidth}{\wideness}} {\advance\wideness by -#1
\ratchet{\leftwidth}{\wideness}} \vertsize{\highness}{#3}
\divide\highness by 2 {\advance\highness by #2
\ratchet{\topheight}{\highness}} {\advance\highness by -#2
\ratchet{\botheight}{\highness}} \put(#1,#2){\makebox(0,0){$#3$}}}

\def\putlbox(#1,#2)#3{%
\horsize{\wideness}{#3} {\advance\wideness by #1
\ratchet{\rightwidth}{\wideness}} {\ratchet{\leftwidth}{-#1}}
\vertsize{\highness}{#3} \divide\highness by 2 {\advance\highness
by #2 \ratchet{\topheight}{\highness}} {\advance\highness by -#2
\ratchet{\botheight}{\highness}}
\put(#1,#2){\makebox(0,0)[l]{$#3$}}}

\def\putrbox(#1,#2)#3{%
\horsize{\wideness}{#3} {\ratchet{\rightwidth}{#1}}
{\advance\wideness by -#1 \ratchet{\leftwidth}{\wideness}}
\vertsize{\highness}{#3} \divide\highness by 2 {\advance\highness
by #2 \ratchet{\topheight}{\highness}} {\advance\highness by -#2
\ratchet{\botheight}{\highness}}
\put(#1,#2){\makebox(0,0)[r]{$#3$}}}

\def\adjust[#1]{} 

\newcount \coefa
\newcount \coefb
\newcount \coefc
\newcount\tempcounta
\newcount\tempcountb
\newcount\tempcountc
\newcount\tempcountd
\newcount\xext
\newcount\yext
\newcount\xoff
\newcount\yoff
\newcount\gap%
\newcount\arrowtypea
\newcount\arrowtypeb
\newcount\arrowtypec
\newcount\arrowtyped
\newcount\arrowtypee
\newcount\height
\newcount\width
\newcount\xpos
\newcount\ypos
\newcount\run
\newcount\rise
\newcount\arrowlength
\newcount\halflength
\newcount\arrowtype
\newdimen\tempdimen
\newdimen\xlen
\newdimen\ylen
\newsavebox{\tempboxa}%
\newsavebox{\tempboxb}%
\newsavebox{\tempboxc}%

\newdimen\w@dth

\def\setw@dth#1#2{\setbox\z@\hbox{\m@th$#1$}\w@dth=\wd\z@
\setbox\@ne\hbox{\m@th$#2$}\ifnum\w@dth<\wd\@ne \w@dth=\wd\@ne \fi
\advance\w@dth by 1.2em}

\def\t@^#1_#2{\allowbreak\def\n@one{#1}\def\n@two{#2}\mathrel
{\setw@dth{#1}{#2} \mathop{\hbox to
\w@dth{\rightarrowfill}}\limits \ifx\n@one\empty\else
^{\box\z@}\fi \ifx\n@two\empty\else _{\box\@ne}\fi}}
\def\t@@^#1{\@ifnextchar_{\t@^{#1}}{\t@^{#1}_{}}}
\def\to{\@ifnextchar^{\t@@}{\t@@^{}}}

\def\t@left^#1_#2{\def\n@one{#1}\def\n@two{#2}\mathrel{\setw@dth{#1}{#2}
\mathop{\hbox to \w@dth{\leftarrowfill}}\limits
\ifx\n@one\empty\else ^{\box\z@}\fi \ifx\n@two\empty\else
_{\box\@ne}\fi}}
\def\t@@left^#1{\@ifnextchar_{\t@left^{#1}}{\t@left^{#1}_{}}}
\def\toleft{\@ifnextchar^{\t@@left}{\t@@left^{}}}

\def\two@^#1_#2{\allowbreak
\def\n@one{#1}\def\n@two{#2}\mathrel{\setw@dth{#1}{#2}
\mathop{\vcenter{\lineskip\z@\baselineskip\z@
                 \hbox to \w@dth{\rightarrowfill}%
                 \hbox to \w@dth{\rightarrowfill}}%
       }\limits
\ifx\n@one\empty\else ^{\box\z@}\fi \ifx\n@two\empty\else
_{\box\@ne}\fi}}
\def\tw@@^#1{\@ifnextchar _{\two@^{#1}}{\two@^{#1}_{}}}
\def\two{\@ifnextchar ^{\tw@@}{\tw@@^{}}}

\def\tofr@^#1_#2{\def\n@one{#1}\def\n@two{#2}\mathrel{\setw@dth{#1}{#2}
\mathop{\vcenter{\hbox to \w@dth{\rightarrowfill}\kern-1.7ex
                 \hbox to \w@dth{\leftarrowfill}}%
       }\limits
\ifx\n@one\empty\else ^{\box\z@}\fi \ifx\n@two\empty\else
_{\box\@ne}\fi}}
\def\t@fr@^#1{\@ifnextchar_ {\tofr@^{#1}}{\tofr@^{#1}_{}}}
\def\tofro{\@ifnextchar^ {\t@fr@}{\t@fr@^{}}}

\def\mon{\mathop{\m@th\hbox to
      14.6\P@{\lasyb\char'51\hskip-2.1\P@$\arrext$\hss
$\mathord\rightarrow$}}\limits} 
\def\leftmono{\mathrel{\m@th\hbox to
14.6\P@{$\mathord\leftarrow$\hss$\arrext$\hskip-2.1\P@\lasyb\char'50%
}}\limits} 
\mathchardef\arrext="0200       

\def\settypes(#1,#2,#3){\arrowtypea#1 \arrowtypeb#2 \arrowtypec#3}
\def\settoheight#1#2{\setbox\@tempboxa\hbox{#2}#1\ht\@tempboxa\relax}%
\def\settodepth#1#2{\setbox\@tempboxa\hbox{#2}#1\dp\@tempboxa\relax}%
\def\settokens`#1`#2`#3`#4`{%
     \def\tokena{#1}\def\tokenb{#2}\def\tokenc{#3}\def\tokend{#4}}
\def\setsqparms[#1`#2`#3`#4;#5`#6]{%
\arrowtypea #1 \arrowtypeb #2 \arrowtypec #3 \arrowtyped #4
\width #5 \height #6 }
\def\setpos(#1,#2){\xpos=#1 \ypos#2}

\def\settriparms[#1`#2`#3;#4]{\settripairparms[#1`#2`#3`1`1;#4]}%

\def\settripairparms[#1`#2`#3`#4`#5;#6]{%
\arrowtypea #1 \arrowtypeb #2 \arrowtypec #3 \arrowtyped #4
\arrowtypee #5 \width #6 \height #6 }

\def\resetparms{\settripairparms[1`1`1`1`1;500]\width 500}

\resetparms

\def\mvector(#1,#2)#3{
\put(0,0){\vector(#1,#2){#3}}%
\put(0,0){\vector(#1,#2){26}}%
}
\def\evector(#1,#2)#3{{
\arrowlength #3
\put(0,0){\vector(#1,#2){\arrowlength}}%
\advance \arrowlength by-30
\put(0,0){\vector(#1,#2){\arrowlength}}%
}}

\def\horsize#1#2{%
\settowidth{\tempdimen}{$#2$}%
#1=\tempdimen \divide #1 by\unitlength }

\def\vertsize#1#2{%
\settoheight{\tempdimen}{$#2$}%
#1=\tempdimen
\settodepth{\tempdimen}{$#2$}%
\advance #1 by\tempdimen \divide #1 by\unitlength }

\def\putvector(#1,#2)(#3,#4)#5#6{{%
\ifnum3<\arrowtype \putdashvector(#1,#2)(#3,#4)#5\arrowtype \else
\ifnum\arrowtype<-3 \putdashvector(#1,#2)(#3,#4)#5\arrowtype \else
\xpos=#1 \ypos=#2 \run=#3 \rise=#4 \arrowlength=#5 \ifnum
\arrowtype<0
    \ifnum \run=0
        \advance \ypos by-\arrowlength
    \else
        \tempcounta \arrowlength
        \multiply \tempcounta by\rise
        \divide \tempcounta by\run
        \ifnum\run>0
            \advance \xpos by\arrowlength
            \advance \ypos by\tempcounta
        \else
            \advance \xpos by-\arrowlength
            \advance \ypos by-\tempcounta
        \fi
    \fi
    \multiply \arrowtype by-1
    \multiply \rise by-1
    \multiply \run by-1
\fi \ifcase \arrowtype
\or \put(\xpos,\ypos){\vector(\run,\rise){\arrowlength}}%
\or \put(\xpos,\ypos){\mvector(\run,\rise)\arrowlength}%
\or \put(\xpos,\ypos){\evector(\run,\rise){\arrowlength}}%
\fi\fi\fi }}

\def\putsplitvector(#1,#2)#3#4{
\xpos #1 \ypos #2 \arrowtype #4 \halflength #3 \arrowlength #3
\gap 140 \advance \halflength by-\gap \divide \halflength by2
\ifnum\arrowtype>0
   \ifcase \arrowtype
   \or \put(\xpos,\ypos){\line(0,-1){\halflength}}%
       \advance\ypos by-\halflength
       \advance\ypos by-\gap
       \put(\xpos,\ypos){\vector(0,-1){\halflength}}%
   \or \put(\xpos,\ypos){\line(0,-1)\halflength}%
       \put(\xpos,\ypos){\vector(0,-1)3}%
       \advance\ypos by-\halflength
       \advance\ypos by-\gap
       \put(\xpos,\ypos){\vector(0,-1){\halflength}}%
   \or \put(\xpos,\ypos){\line(0,-1)\halflength}%
       \advance\ypos by-\halflength
       \advance\ypos by-\gap
       \put(\xpos,\ypos){\evector(0,-1){\halflength}}%
   \fi
\else \arrowtype=-\arrowtype
   \ifcase\arrowtype
   \or \advance \ypos by-\arrowlength
       \put(\xpos,\ypos){\line(0,1){\halflength}}%
       \advance\ypos by\halflength
       \advance\ypos by\gap
       \put(\xpos,\ypos){\vector(0,1){\halflength}}%
   \or \advance \ypos by-\arrowlength
       \put(\xpos,\ypos){\line(0,1)\halflength}%
       \put(\xpos,\ypos){\vector(0,1)3}%
       \advance\ypos by\halflength
       \advance\ypos by\gap
       \put(\xpos,\ypos){\vector(0,1){\halflength}}%
   \or \advance \ypos by-\arrowlength
       \put(\xpos,\ypos){\line(0,1)\halflength}%
       \advance\ypos by\halflength
       \advance\ypos by\gap
       \put(\xpos,\ypos){\evector(0,1){\halflength}}%
   \fi
\fi }

\def\putmorphism(#1)(#2,#3)[#4`#5`#6]#7#8#9{{%
\run #2 \rise #3 \ifnum\rise=0
  \puthmorphism(#1)[#4`#5`#6]{#7}{#8}#9%
\else\ifnum\run=0
  \putvmorphism(#1)[#4`#5`#6]{#7}{#8}#9%
\else
\setpos(#1)%
\arrowlength #7 \arrowtype #8 \ifnum\run=0 \else\ifnum\rise=0
\else \ifnum\run>0
    \coefa=1
\else
   \coefa=-1
\fi \ifnum\arrowtype>0
   \coefb=0
   \coefc=-1
\else
   \coefb=\coefa
   \coefc=1
   \arrowtype=-\arrowtype
\fi \width=2 \multiply \width by\run \divide \width by\rise
\ifnum \width<0  \width=-\width\fi \advance\width by60 \if l#9
\width=-\width\fi
\putbox(\xpos,\ypos){#4}
{\multiply \coefa by\arrowlength
\advance\xpos by\coefa \multiply \coefa by\rise \divide \coefa
by\run \advance \ypos by\coefa
\putbox(\xpos,\ypos){#5} }%
{\multiply \coefa by\arrowlength
\divide \coefa by2 \advance \xpos by\coefa \advance \xpos by\width
\multiply \coefa by\rise \divide \coefa by\run \advance \ypos
by\coefa
\if l#9%
   \putrbox(\xpos,\ypos){#6}%
\else\if r#9%
   \putlbox(\xpos,\ypos){#6}%
\fi\fi }%
{\multiply \rise by-\coefc
\multiply \run by-\coefc \multiply \coefb by\arrowlength \advance
\xpos by\coefb \multiply \coefb by\rise \divide \coefb by\run
\advance \ypos by\coefb \multiply \coefc by70 \advance \ypos
by\coefc \multiply \coefc by\run \divide \coefc by\rise \advance
\xpos by\coefc \multiply \coefa by140 \multiply \coefa by\run
\divide \coefa by\rise \advance \arrowlength by\coefa
\ifcase\arrowtype
\or \put(\xpos,\ypos){\vector(\run,\rise){\arrowlength}}%
\or \put(\xpos,\ypos){\mvector(\run,\rise){\arrowlength}}%
\or \put(\xpos,\ypos){\evector(\run,\rise){\arrowlength}}%
\fi}\fi\fi\fi\fi}}

\newcount\numbdashes \newcount\lengthdash \newcount\increment

\def\howmanydashes{
\numbdashes=\arrowlength \lengthdash=40 \divide\numbdashes by
\lengthdash \lengthdash=\arrowlength \divide\lengthdash by
\numbdashes
\increment=\lengthdash \multiply\lengthdash by 3
\divide\lengthdash by 5 }

\def\putdashvector(#1)(#2,#3)#4#5{%
\ifnum#3=0 \putdashhvector(#1){#4}#5 \else \ifnum#2=0
\putdashvvector(#1){#4}#5\fi\fi}

\def\putdashhvector(#1,#2)#3#4{{%
\arrowlength=#3 \howmanydashes
\multiput(#1,#2)(\increment,0){\numbdashes}%
{\vrule height .4pt width \lengthdash\unitlength} \arrowtype=#4
\xpos=#1 \ifnum\arrowtype<0 \advance\arrowtype by 7 \fi
\ifcase\arrowtype \or \advance\xpos by 10
    \put(\xpos,#2){\vector(-1,0){\lengthdash}}
    \advance\xpos by 40
    \put(\xpos,#2){\vector(-1,0){\lengthdash}}
\or \advance \xpos by 10
    \put(\xpos,#2){\vector(-1,0){\lengthdash}}
    \advance\xpos by  \arrowlength
    \advance\xpos by  -50
    \put(\xpos,#2){\vector(-1,0){\lengthdash}}
\or \advance\xpos by 10
    \put(\xpos,#2){\vector(-1,0){\lengthdash}}
\or \advance\xpos by \arrowlength
    \advance\xpos by -\lengthdash
    \put(\xpos,#2){\vector(1,0){\lengthdash}}
\or {\advance\xpos by 10
    \put(\xpos,#2){\vector(1,0){\lengthdash}}}
    \advance\xpos by \arrowlength
    \advance\xpos by -\lengthdash
    \put(\xpos,#2){\vector(1,0){\lengthdash}}
\or \advance\xpos by \arrowlength
    \advance\xpos by -\lengthdash
    \put(\xpos,#2){\vector(1,0){\lengthdash}}
    \advance\xpos by -40
    \put(\xpos,#2){\vector(1,0){\lengthdash}}
   \fi
}}

\def\putdashvvector(#1,#2)#3#4{{%
\arrowlength=#3 \howmanydashes \ypos=#2 \advance\ypos by
-\arrowlength
\multiput(#1,#2)(0,\increment){\numbdashes}%
    {\vrule width .4pt height \lengthdash\unitlength}
\arrowtype=#4 \ypos=#2 \ifnum\arrowtype<0 \advance\arrowtype by 7
\fi \ifcase\arrowtype \or \advance\ypos by \arrowlength
\advance\ypos by -40
    \put(#1,\ypos){\vector(0,1){\lengthdash}}
    \advance\ypos by -40
    \put(#1,\ypos){\vector(0,1){\lengthdash}}
\or \advance\ypos by 10
    \put(#1,\ypos){\vector(0,1){\lengthdash}}
    \advance\ypos by \arrowlength \advance\ypos by -40
    \put(#1,\ypos){\vector(0,1){\lengthdash}}
\or \advance\ypos by \arrowlength \advance\ypos by -40
    \put(#1,\ypos){\vector(0,1){\lengthdash}}
\or \advance\ypos by 10
    \put(#1,\ypos){\vector(0,-1){\lengthdash}}
\or \advance\ypos by 10
    \put(#1,\ypos){\vector(0,-1){\lengthdash}}
    \advance\ypos by \arrowlength \advance\ypos by -40
    \put(#1,\ypos){\vector(0,-1){\lengthdash}}
\or \advance\ypos by 10
    \put(#1,\ypos){\vector(0,-1){\lengthdash}}
    \advance\ypos by 40
    \put(#1,\ypos){\vector(0,-1){\lengthdash}}
\fi }}

\def\puthmorphism(#1,#2)[#3`#4`#5]#6#7#8{{%
\xpos #1 \ypos #2 \width #6 \arrowlength #6 \arrowtype=#7
\putbox(\xpos,\ypos){#3\vphantom{#4}}%
{\advance \xpos by\arrowlength
\putbox(\xpos,\ypos){\vphantom{#3}#4}}%
\horsize{\tempcounta}{#3}%
\horsize{\tempcountb}{#4}%
\divide \tempcounta by2 \divide \tempcountb by2 \advance
\tempcounta by30 \advance \tempcountb by30 \advance \xpos
by\tempcounta \advance \arrowlength by-\tempcounta \advance
\arrowlength by-\tempcountb
\putvector(\xpos,\ypos)(1,0)\arrowlength\arrowtype \divide
\arrowlength by2 \advance \xpos by\arrowlength
\vertsize{\tempcounta}{#5}%
\divide\tempcounta by2 \advance \tempcounta by20
\if a#8 %
   \advance \ypos by\tempcounta
   \putbox(\xpos,\ypos){#5}%
\else
   \advance \ypos by-\tempcounta
   \putbox(\xpos,\ypos){#5}%
\fi}}

\def\putvmorphism(#1,#2)[#3`#4`#5]#6#7#8{{%
\xpos #1 \ypos #2 \arrowlength #6 \arrowtype #7
\settowidth{\xlen}{$#5$}%
\putbox(\xpos,\ypos){#3}%
{\advance \ypos by-\arrowlength
\putbox(\xpos,\ypos){#4}}%
{\advance\arrowlength by-140 \advance \ypos by-70 \ifdim\xlen>0pt
   \if m#8%
      \putsplitvector(\xpos,\ypos)\arrowlength\arrowtype
   \else
   \putvector(\xpos,\ypos)(0,-1)\arrowlength\arrowtype
   \fi
\else
   \putvector(\xpos,\ypos)(0,-1)\arrowlength\arrowtype
\fi}%
\ifdim\xlen>0pt
   \divide \arrowlength by2
   \advance\ypos by-\arrowlength
   \if l#8%
      \advance \xpos by-40
      \putrbox(\xpos,\ypos){#5}%
   \else\if r#8%
      \advance \xpos by40
      \putlbox(\xpos,\ypos){#5}%
   \else
      \putbox(\xpos,\ypos){#5}%
   \fi\fi
\fi }}

\def\putsquarep<#1>(#2)[#3;#4`#5`#6`#7]{{%
\setsqparms[#1]%
\setpos(#2)%
\settokens`#3`%
\puthmorphism(\xpos,\ypos)[\tokenc`\tokend`{#7}]{\width}{\arrowtyped}b%
\advance\ypos by \height
\puthmorphism(\xpos,\ypos)[\tokena`\tokenb`{#4}]{\width}{\arrowtypea}a%
\putvmorphism(\xpos,\ypos)[``{#5}]{\height}{\arrowtypeb}l%
\advance\xpos by \width
\putvmorphism(\xpos,\ypos)[``{#6}]{\height}{\arrowtypec}r%
}}

\def\putsquare{\@ifnextchar <{\putsquarep}{\putsquarep%
   <\arrowtypea`\arrowtypeb`\arrowtypec`\arrowtyped;\width`\height>}}
\def\square{\@ifnextchar< {\squarep}{\squarep
   <\arrowtypea`\arrowtypeb`\arrowtypec`\arrowtyped;\width`\height>}}
\def\squarep<#1>[#2`#3`#4`#5;#6`#7`#8`#9]{{
\setsqparms[#1]
\diagram
\putsquarep<\arrowtypea`\arrowtypeb`\arrowtypec`
\arrowtyped;\width`\height>
(0,0)[#2`#3`#4`{#5};#6`#7`#8`{#9}]
\enddiagram
}}                                                 
\def\putptrianglep<#1>(#2,#3)[#4`#5`#6;#7`#8`#9]{{%
\settriparms[#1]%
\xpos=#2 \ypos=#3 \advance\ypos by \height
\puthmorphism(\xpos,\ypos)[#4`#5`{#7}]{\height}{\arrowtypea}a%
\putvmorphism(\xpos,\ypos)[`#6`{#8}]{\height}{\arrowtypeb}l%
\advance\xpos by\height
\putmorphism(\xpos,\ypos)(-1,-1)[``{#9}]{\height}{\arrowtypec}r%
}}

\def\putptriangle{\@ifnextchar <{\putptrianglep}{\putptrianglep
   <\arrowtypea`\arrowtypeb`\arrowtypec;\height>}}
\def\ptriangle{\@ifnextchar <{\ptrianglep}{\ptrianglep
   <\arrowtypea`\arrowtypeb`\arrowtypec;\height>}}
\def\ptrianglep<#1>[#2`#3`#4;#5`#6`#7]{{
\settriparms[#1]
\diagram
\putptrianglep<\arrowtypea`\arrowtypeb`
\arrowtypec;\height>
(0,0)[#2`#3`#4;#5`#6`{#7}]
\enddiagram
}}                                            

\def\putqtrianglep<#1>(#2,#3)[#4`#5`#6;#7`#8`#9]{{%
\settriparms[#1]%
\xpos=#2 \ypos=#3 \advance\ypos by\height
\puthmorphism(\xpos,\ypos)[#4`#5`{#7}]{\height}{\arrowtypea}a%
\putmorphism(\xpos,\ypos)(1,-1)[``{#8}]{\height}{\arrowtypeb}l%
\advance\xpos by\height
\putvmorphism(\xpos,\ypos)[`#6`{#9}]{\height}{\arrowtypec}r%
}}

\def\putqtriangle{\@ifnextchar <{\putqtrianglep}{\putqtrianglep
   <\arrowtypea`\arrowtypeb`\arrowtypec;\height>}}
\def\qtriangle{\@ifnextchar <{\qtrianglep}{\qtrianglep
   <\arrowtypea`\arrowtypeb`\arrowtypec;\height>}}
\def\qtrianglep<#1>[#2`#3`#4;#5`#6`#7]{{
\settriparms[#1]
\width=\height                                
\diagram
\putqtrianglep<\arrowtypea`\arrowtypeb`
\arrowtypec;\height>
(0,0)[#2`#3`#4;#5`#6`{#7}]
\enddiagram
}}

\def\putdtrianglep<#1>(#2,#3)[#4`#5`#6;#7`#8`#9]{{%
\settriparms[#1]%
\xpos=#2 \ypos=#3
\puthmorphism(\xpos,\ypos)[#5`#6`{#9}]{\height}{\arrowtypec}b%
\advance\xpos by \height \advance\ypos by\height
\putmorphism(\xpos,\ypos)(-1,-1)[``{#7}]{\height}{\arrowtypea}l%
\putvmorphism(\xpos,\ypos)[#4``{#8}]{\height}{\arrowtypeb}r%
}}

\def\putdtriangle{\@ifnextchar <{\putdtrianglep}{\putdtrianglep
   <\arrowtypea`\arrowtypeb`\arrowtypec;\height>}}
\def\dtriangle{\@ifnextchar <{\dtrianglep}{\dtrianglep
   <\arrowtypea`\arrowtypeb`\arrowtypec;\height>}}
\def\dtrianglep<#1>[#2`#3`#4;#5`#6`#7]{{
\settriparms[#1]
\width=\height                                
\diagram
\putdtrianglep<\arrowtypea`\arrowtypeb`
\arrowtypec;\height>
(0,0)[#2`#3`#4;#5`#6`{#7}]
\enddiagram
}}

\def\putbtrianglep<#1>(#2,#3)[#4`#5`#6;#7`#8`#9]{{%
\settriparms[#1]%
\xpos=#2 \ypos=#3
\puthmorphism(\xpos,\ypos)[#5`#6`{#9}]{\height}{\arrowtypec}b%
\advance\ypos by\height
\putmorphism(\xpos,\ypos)(1,-1)[``{#8}]{\height}{\arrowtypeb}r%
\putvmorphism(\xpos,\ypos)[#4``{#7}]{\height}{\arrowtypea}l%
}}

\def\putbtriangle{\@ifnextchar <{\putbtrianglep}{\putbtrianglep
   <\arrowtypea`\arrowtypeb`\arrowtypec;\height>}}
\def\btriangle{\@ifnextchar <{\btrianglep}{\btrianglep
   <\arrowtypea`\arrowtypeb`\arrowtypec;\height>}}
\def\btrianglep<#1>[#2`#3`#4;#5`#6`#7]{{
\settriparms[#1]
\width=\height                               
\diagram
\putbtrianglep<\arrowtypea`\arrowtypeb`
\arrowtypec;\height>
(0,0)[#2`#3`#4;#5`#6`{#7}]
\enddiagram
}}

\def\putAtrianglep<#1>(#2,#3)[#4`#5`#6;#7`#8`#9]{{%
\settriparms[#1]%
\xpos=#2 \ypos=#3 {\multiply \height by2
\puthmorphism(\xpos,\ypos)[#5`#6`{#9}]{\height}{\arrowtypec}b}%
\advance\xpos by\height \advance\ypos by\height
\putmorphism(\xpos,\ypos)(-1,-1)[#4``{#7}]{\height}{\arrowtypea}l%
\putmorphism(\xpos,\ypos)(1,-1)[``{#8}]{\height}{\arrowtypeb}r%
}}

\def\putAtriangle{\@ifnextchar <{\putAtrianglep}{\putAtrianglep
   <\arrowtypea`\arrowtypeb`\arrowtypec;\height>}}
\def\Atriangle{\@ifnextchar <{\Atrianglep}{\Atrianglep
   <\arrowtypea`\arrowtypeb`\arrowtypec;\height>}}
\def\Atrianglep<#1>[#2`#3`#4;#5`#6`#7]{{
\settriparms[#1]
\width=\height                                     
\diagram
\putAtrianglep<\arrowtypea`\arrowtypeb`
\arrowtypec;\height>
(0,0)[#2`#3`#4;#5`#6`{#7}]
\enddiagram
}}

\def\putAtrianglepairp<#1>(#2)[#3;#4`#5`#6`#7`#8]{{%
\settripairparms[#1]%
\setpos(#2)%
\settokens`#3`%
\puthmorphism(\xpos,\ypos)[\tokenb`\tokenc`{#7}]{\height}{\arrowtyped}b%
\advance\xpos by\height
\puthmorphism(\xpos,\ypos)[\phantom{\tokenc}`\tokend`{#8}]%
{\height}{\arrowtypee}b%
\advance\ypos by\height
\putmorphism(\xpos,\ypos)(-1,-1)[\tokena``{#4}]{\height}{\arrowtypea}l%
\putvmorphism(\xpos,\ypos)[``{#5}]{\height}{\arrowtypeb}m%
\putmorphism(\xpos,\ypos)(1,-1)[``{#6}]{\height}{\arrowtypec}r%
}}

\def\putAtrianglepair{\@ifnextchar <{\putAtrianglepairp}{\putAtrianglepairp%
   <\arrowtypea`\arrowtypeb`\arrowtypec`\arrowtyped`\arrowtypee;\height>}}
\def\Atrianglepair{\@ifnextchar <{\Atrianglepairp}{\Atrianglepairp%
   <\arrowtypea`\arrowtypeb`\arrowtypec`\arrowtyped`\arrowtypee;\height>}}

\def\Atrianglepairp<#1>[#2;#3`#4`#5`#6`#7]{{
\settripairparms[#1]
\settokens`#2`
\width=\height                                
\diagram
\putAtrianglepairp                            
<\arrowtypea`\arrowtypeb`\arrowtypec`
\arrowtyped`\arrowtypee;\height>
(0,0)[{#2};#3`#4`#5`#6`{#7}]
\enddiagram
}}

\def\putVtrianglep<#1>(#2,#3)[#4`#5`#6;#7`#8`#9]{{%
\settriparms[#1]%
\xpos=#2 \ypos=#3 \advance\ypos by\height {\multiply\height by2
\puthmorphism(\xpos,\ypos)[#4`#5`{#7}]{\height}{\arrowtypea}a}%
\putmorphism(\xpos,\ypos)(1,-1)[`#6`{#8}]{\height}{\arrowtypeb}l%
\advance\xpos by\height \advance\xpos by\height
\putmorphism(\xpos,\ypos)(-1,-1)[``{#9}]{\height}{\arrowtypec}r%
}}

\def\putVtriangle{\@ifnextchar <{\putVtrianglep}{\putVtrianglep
   <\arrowtypea`\arrowtypeb`\arrowtypec;\height>}}
\def\Vtriangle{\@ifnextchar <{\Vtrianglep}{\Vtrianglep
   <\arrowtypea`\arrowtypeb`\arrowtypec;\height>}}
\def\Vtrianglep<#1>[#2`#3`#4;#5`#6`#7]{{
\settriparms[#1]
\width=\height                                 
\diagram
\putVtrianglep<\arrowtypea`\arrowtypeb`
\arrowtypec;\height>
(0,0)[#2`#3`#4;#5`#6`{#7}]
\enddiagram
}}

\def\putVtrianglepairp<#1>(#2)[#3;#4`#5`#6`#7`#8]{{
\settripairparms[#1]%
\setpos(#2)%
\settokens`#3`%
\advance\ypos by\height
\putmorphism(\xpos,\ypos)(1,-1)[`\tokend`{#6}]{\height}{\arrowtypec}l%
\puthmorphism(\xpos,\ypos)[\tokena`\tokenb`{#4}]{\height}{\arrowtypea}a%
\advance\xpos by\height
\puthmorphism(\xpos,\ypos)[\phantom{\tokenb}`\tokenc`{#5}]%
{\height}{\arrowtypeb}a%
\putvmorphism(\xpos,\ypos)[``{#7}]{\height}{\arrowtyped}m%
\advance\xpos by\height
\putmorphism(\xpos,\ypos)(-1,-1)[``{#8}]{\height}{\arrowtypee}r%
}}

\def\putVtrianglepair{\@ifnextchar <{\putVtrianglepairp}{\putVtrianglepairp%
    <\arrowtypea`\arrowtypeb`\arrowtypec`\arrowtyped`\arrowtypee;\height>}}
\def\Vtrianglepair{\@ifnextchar <{\Vtrianglepairp}{\Vtrianglepairp%
    <\arrowtypea`\arrowtypeb`\arrowtypec`\arrowtyped`\arrowtypee;\height>}}
\def\Vtrianglepairp<#1>[#2;#3`#4`#5`#6`#7]{{
\settripairparms[#1]
\settokens`#2`
\diagram
\putVtrianglepairp                             
<\arrowtypea`\arrowtypeb`\arrowtypec`
\arrowtyped`\arrowtypee;\height>
(0,0)[{#2};#3`#4`#5`#6`{#7}]
\enddiagram
}}

\def\putCtrianglep<#1>(#2,#3)[#4`#5`#6;#7`#8`#9]{{%
\settriparms[#1]%
\xpos=#2 \ypos=#3 \advance\ypos by\height
\putmorphism(\xpos,\ypos)(1,-1)[``{#9}]{\height}{\arrowtypec}l%
\advance\xpos by\height \advance\ypos by\height
\putmorphism(\xpos,\ypos)(-1,-1)[#4`#5`{#7}]{\height}{\arrowtypea}l%
{\multiply\height by 2
\putvmorphism(\xpos,\ypos)[`#6`{#8}]{\height}{\arrowtypeb}r}%
}}

\def\putCtriangle{\@ifnextchar <{\putCtrianglep}{\putCtrianglep
    <\arrowtypea`\arrowtypeb`\arrowtypec;\height>}}
\def\Ctriangle{\@ifnextchar <{\Ctrianglep}{\Ctrianglep
    <\arrowtypea`\arrowtypeb`\arrowtypec;\height>}}
\def\Ctrianglep<#1>[#2`#3`#4;#5`#6`#7]{{
\settriparms[#1]
\width=\height                               
\diagram
\putCtrianglep<\arrowtypea`\arrowtypeb`
\arrowtypec;\height>
(0,0)[#2`#3`#4;#5`#6`{#7}]
\enddiagram
}}                                           
\def\putDtrianglep<#1>(#2,#3)[#4`#5`#6;#7`#8`#9]{{%
\settriparms[#1]%
\xpos=#2 \ypos=#3 \advance\xpos by\height \advance\ypos by\height
\putmorphism(\xpos,\ypos)(-1,-1)[``{#9}]{\height}{\arrowtypec}r%
\advance\xpos by-\height \advance\ypos by\height
\putmorphism(\xpos,\ypos)(1,-1)[`#5`{#8}]{\height}{\arrowtypeb}r%
{\multiply\height by 2
\putvmorphism(\xpos,\ypos)[#4`#6`{#7}]{\height}{\arrowtypea}l}%
}}

\def\putDtriangle{\@ifnextchar <{\putDtrianglep}{\putDtrianglep
    <\arrowtypea`\arrowtypeb`\arrowtypec;\height>}}
\def\Dtriangle{\@ifnextchar <{\Dtrianglep}{\Dtrianglep
   <\arrowtypea`\arrowtypeb`\arrowtypec;\height>}}
\def\Dtrianglep<#1>[#2`#3`#4;#5`#6`#7]{{
\settriparms[#1]
\width=\height                              
\diagram
\putDtrianglep<\arrowtypea`\arrowtypeb`
\arrowtypec;\height>
(0,0)[#2`#3`#4;#5`#6`{#7}]
\enddiagram
}}                                          
\def\setrecparms[#1`#2]{\width=#1 \height=#2}%

\def\recursep<#1`#2>[#3;#4`#5`#6`#7`#8]{{\m@th
\width=#1 \height=#2 \settokens`#3`
\settowidth{\tempdimen}{$\tokena$} \ifdim\tempdimen=0pt
  \savebox{\tempboxa}{\hbox{$\tokenb$}}%
  \savebox{\tempboxb}{\hbox{$\tokend$}}%
  \savebox{\tempboxc}{\hbox{$#6$}}%
\else
  \savebox{\tempboxa}{\hbox{$\hbox{$\tokena$}\times\hbox{$\tokenb$}$}}%
  \savebox{\tempboxb}{\hbox{$\hbox{$\tokena$}\times\hbox{$\tokend$}$}}%
  \savebox{\tempboxc}{\hbox{$\hbox{$\tokena$}\times\hbox{$#6$}$}}%
\fi \ypos=\height \divide\ypos by 2 \xpos=\ypos \advance\xpos by
\width \bfig
\putCtrianglep<-1`1`1;\ypos>(0,0)[`\tokenc`;#5`#6`{#7}]%
\puthmorphism(\ypos,0)[\tokend`\usebox{\tempboxb}`{#8}]{\width}{-1}b%
\puthmorphism(\ypos,\height)[\tokenb`\usebox{\tempboxa}`{#4}]{\width}{-1}a%
\advance\ypos by \width
\putvmorphism(\ypos,\height)[``\usebox{\tempboxc}]{\height}1r%
\efig }}

\def\recurse{\@ifnextchar <{\recursep}{\recursep<\width`\height>}}

\def\puttwohmorphisms(#1,#2)[#3`#4;#5`#6]#7#8#9{{%
\puthmorphism(#1,#2)[#3`#4`]{#7}0a \ypos=#2 \advance\ypos by 20
\puthmorphism(#1,\ypos)[\phantom{#3}`\phantom{#4}`#5]{#7}{#8}a
\advance\ypos by -40
\puthmorphism(#1,\ypos)[\phantom{#3}`\phantom{#4}`#6]{#7}{#9}b }}

\def\puttwovmorphisms(#1,#2)[#3`#4;#5`#6]#7#8#9{{%
\putvmorphism(#1,#2)[#3`#4`]{#7}0a \xpos=#1 \advance\xpos by -20
\putvmorphism(\xpos,#2)[\phantom{#3}`\phantom{#4}`#5]{#7}{#8}l
\advance\xpos by 40
\putvmorphism(\xpos,#2)[\phantom{#3}`\phantom{#4}`#6]{#7}{#9}r }}

\def\puthcoequalizer(#1)[#2`#3`#4;#5`#6`#7]#8#9{{%
\setpos(#1)%
\puttwohmorphisms(\xpos,\ypos)[#2`#3;#5`#6]{#8}11%
\advance\xpos by #8
\puthmorphism(\xpos,\ypos)[\phantom{#3}`#4`#7]{#8}1{#9} }}

\def\putvcoequalizer(#1)[#2`#3`#4;#5`#6`#7]#8#9{{%
\setpos(#1)%
\puttwovmorphisms(\xpos,\ypos)[#2`#3;#5`#6]{#8}11%
\advance\ypos by -#8
\putvmorphism(\xpos,\ypos)[\phantom{#3}`#4`#7]{#8}1{#9} }}

\def\putthreehmorphisms(#1)[#2`#3;#4`#5`#6]#7(#8)#9{{%
\setpos(#1) \settypes(#8)
\if a#9 %
     \vertsize{\tempcounta}{#5}%
     \vertsize{\tempcountb}{#6}%
     \ifnum \tempcounta<\tempcountb \tempcounta=\tempcountb \fi
\else
     \vertsize{\tempcounta}{#4}%
     \vertsize{\tempcountb}{#5}%
     \ifnum \tempcounta<\tempcountb \tempcounta=\tempcountb \fi
\fi \advance \tempcounta by 60
\puthmorphism(\xpos,\ypos)[#2`#3`#5]{#7}{\arrowtypeb}{#9}
\advance\ypos by \tempcounta
\puthmorphism(\xpos,\ypos)[\phantom{#2}`\phantom{#3}`#4]{#7}{\arrowtypea}{#9}
\advance\ypos by -\tempcounta \advance\ypos by -\tempcounta
\puthmorphism(\xpos,\ypos)[\phantom{#2}`\phantom{#3}`#6]{#7}{\arrowtypec}{#9}
}}

\def\setarrowtoks[#1`#2`#3`#4`#5`#6]{%
\def\toka{#1}
\def\tokb{#2}
\def\tokc{#3}
\def\tokd{#4}
\def\toke{#5}
\def\tokf{#6}
}
\def\hex{\@ifnextchar <{\hexp}{\hexp<1000`400>}}
\def\hexp<#1`#2>[#3`#4`#5`#6`#7`#8;#9]{%
\setarrowtoks[#9] \yext=#2 \advance \yext by #2 \xext=#1
\advance\xext by \yext \bfig
\putCtriangle<-1`0`1;#2>(0,0)[`#5`;\tokb``\tokd] \xext=#1
\yext=#2 \advance \yext by #2
\putsquare<1`0`0`1;\xext`\yext>(#2,0)[#3`#4`#7`#8;\toka```\tokf]
\advance \xext by #2
\putDtriangle<0`1`-1;#2>(\xext,0)[`#6`;`\tokc`\toke] \efig }

\makeatother

\input{tcilatex}

\newcommand{\la}{\lambda}
\newcommand{\al}{\alpha}

\newcommand{\w}{\wedge}

\newcommand{\m}{\mu}

\newcommand{\G}{\Gamma}
\newcommand{\beq}{\begin{equation}}
\newcommand{\eeq}{\end{equation}}
\newcommand{\be}{\begin{eqnarray*}}
\newcommand{\ee}{\end{eqnarray*}}

\def\op#1{\mathop{\fam0 #1}\limits}

\begin{document}

\title{Jet Spaces in Modern Hamiltonian Biomechanics}
\author{Tijana T. Ivancevic,\thanks{Citech Research IP Pty Ltd. \& QLIWW IP Pty Ltd., Adelaide, Australia} Bojan Jovanovic,\thanks{University of Nis, Serbia}
Ratko Stankovic,\thanks{Ibid} and Sasa Markovic\thanks{Ibid}}
\date{}
\maketitle

\begin{abstract}
In this paper we propose the time-dependent Hamiltonian form of
human biomechanics, as a sequel to our previous work in time-dependent Lagrangian biomechanics \cite{TijLagr}. Starting with the Covariant Force Law, we first develop
autonomous Hamiltonian biomechanics. Then we extend it using a powerful
geometrical machinery consisting of fibre bundles and jet manifolds associated to the biomechanical configuration manifold. We derive
time-dependent, dissipative, Hamiltonian equations and the fitness evolution
equation for the general time-dependent human biomechanical
system.\newline

\end{abstract}


\section{Introduction}

Most of dynamics in contemporary human biomechanics is \emph{autonomous}
(see \cite
{GaneshSprSml,GaneshWSc,GaneshSprBig,GaneshADG,StrAttr,TijIJHR,Complexity,TijNis,TijNL,TijSpr}).
This approach works fine for most individual movement simulations and
predictions, in which the total human energy dissipations are insignificant.
However, if we analyze a 100\thinspace m-dash sprinting motion, which is in
case of top athletes finished under 10\thinspace s, we can recognize a
significant slow-down after about 70\thinspace m in \emph{all} athletes --
despite of their strong intention to finish and win the race, which is an
obvious sign of the total energy dissipation. This can be seen, for example,
in a current record-braking speed-distance curve of Usain Bolt, the
world-record holder with 9.69\ s, or in a former record-braking
speed--distance curve of Carl Lewis, the former world-record holder (and 9
time Olympic gold medalist) with 9.86\ s (see Figure 3.7 in \cite{TijSpr}).
In other words, the \emph{total mechanical energy} of a sprinter \emph{%
cannot be conserved} even for 10\thinspace s. So, if we want to develop a
realistic model of intensive human motion that is longer than 7--8\thinspace
s, we necessarily need to use the more advanced formalism of time-dependent
mechanics.

In this paper, as a sequel to our previous work in time-dependent Lagrangian biomechanics \cite{TijLagr}, we use the covariant force law in conjunction with the modern geometric formalism of jet manifolds to develop
general Hamiltonian approach to time-dependent human biomechanics in which
\emph{total mechanical energy is not conserved}.

\section{The Covariant Force Law}

Autonomous Hamiltonian biomechanics (as well as autonomous Lagrangian
biomechanics), based on the postulate of conservation of the total
mechanical energy, can be derived from the \textit{covariant force law} \cite
{GaneshSprSml,GaneshWSc,GaneshSprBig,GaneshADG}, which in `plain English'
states:
\[
\text{Force 1--form}=\text{Mass distribution}\times \text{Acceleration
vector-field},
\]
and formally reads (using Einstein's summation convention over repeated
indices):
\begin{equation}
F_{i}=m_{ij}a^{j}.  \label{ivcov}
\end{equation}
Here, the force 1--form $F_{i}=F_{i}(t,q,p)=F^{\prime}_{i}(t,q,\dot{q}%
),~(i=1,...,n)$ denotes any type of torques and forces acting on a human
skeleton, including excitation and contraction dynamics of
muscular--actuators \cite{Hill,Wilkie,Hatze} and rotational dynamics of
hybrid robot actuators, as well as (nonlinear) dissipative joint torques and
forces and external stochastic perturbation torques and forces \cite{StrAttr}%
. $m_{ij}$ is the material (mass--inertia) metric tensor, which gives the
total mass distribution of the human body, by including all segmental masses
and their individual inertia tensors. $a^{j}$ is the total acceleration
vector-field, including all segmental vector-fields, defined as the absolute
(Bianchi) derivative $\dot{\bar v}^i$ of all the segmental angular and
linear velocities $v^i=\dot{x}^i,~(i=1,...,n)$, where $n$ is the total
number of active degrees of freedom (DOF) with local coordinates $(x^i)$.
\begin{figure}[h]
\centerline{\includegraphics[width=10cm]{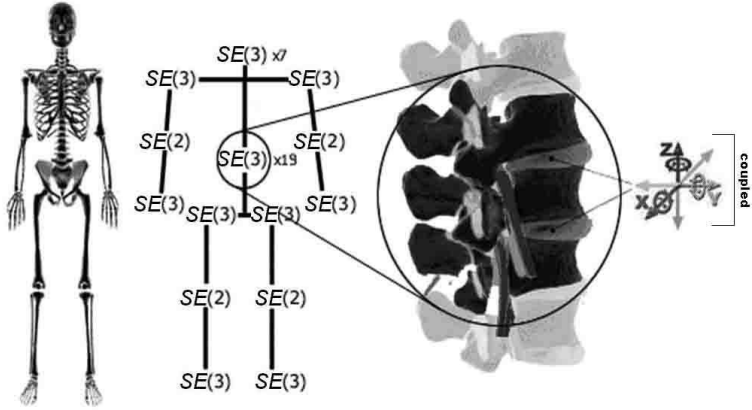}}
\caption{Biomechanical configuration manifold defined as an anthropomorphic product tree consisting of Euclidean motion groups SE(3) (described in the text below). }
\label{SpineSE3}
\end{figure}

More formally, this \emph{central Law of biomechanics} represents the
\textit{covariant force functor} $\mathcal{F}_*$ defined by the commutative
diagram:

\begin{equation}
{\large \putsquare<1`-1`-1`0;1100`500>(360,500)[TT^*M`TTM`` ;\mathcal{F}%
_*`F_i= \dot{p}_i`a^i= \dot{\bar{v}}^i`] \Vtriangle<0`-1`-1;>[T^*M=
\{x^i,p_i\}\;\;`\;\;TM= \{x^i,v^i\}`\;\;M= \{x^i\};`p_i `v^i= \dot{x}^i] }
\label{covfun}
\end{equation}
\smallskip

Here, $M\equiv M^n=\{x^i,~(i=1,...,n)\}$ is the biomechanical configuration $%
n-$manifold (Figure \ref{SpineSE3}), that is the set of all active DOF of the biomechanical system
under consideration (in general, human skeleton), with local coordinates $%
(x^i)$.

The right-hand branch of the fundamental covariant force functor $\mathcal{F}%
_*:TT^*M \to TTM$ depicted in (\ref{covfun}) is Lagrangian dynamics with its
Riemannian geometry. To each $n-$dimensional ($n$D) smooth manifold $M$
there is associated its $2n$D \textit{velocity phase-space manifold},
denoted by $TM$ and called the tangent bundle of $M$. The original
configuration manifold $M$ is called the \textit{base} of $TM$. There is an
onto map $\pi :TM\rightarrow M$, called the \textit{projection}. Above each
point $x\in M$ there is a {tangent space} $T_{x}M=\pi ^{-1}(x)$ to $M$ at $x$%
, which is called a {fibre}. The fibre $T_{x}M\subset TM$ is the subset of $%
TM $, such that the total tangent bundle, $TM=\dbigsqcup\limits_{m\in
M}T_{x}M$, is a {disjoint union} of tangent spaces $T_{x}M$ to $M$ for all
points $x\in M$. From dynamical perspective, the most important quantity in
the tangent bundle concept is the smooth map $v:M\rightarrow TM$, which is
an inverse to the projection $\pi $, i.e, $\pi \circ v=\func{Id}_{M},\;\pi
(v(x))=x$. It is called the \textit{velocity vector-field} $v^i= \dot{x}^i$.%
\footnote{%
This explains the dynamical term \textit{velocity phase--space}, given to
the tangent bundle $TM$ of the manifold $M$.} Its graph $(x,v(x))$
represents the {cross--section} of the tangent bundle $TM$. Velocity
vector-fields are cross-sections of the tangent bundle. Biomechanical \emph{%
Lagrangian} (that is, kinetic minus potential energy) is a natural energy
function on the tangent bundle $TM$. The tangent bundle is itself a smooth
manifold. It has its own tangent bundle, $TTM$. Cross-sections of the second
tangent bundle $TTM$ are the acceleration vector-fields.

The left-hand branch of the fundamental covariant force functor $\mathcal{F}%
_*:TT^*M \to TTM$ depicted in (\ref{covfun}) is Hamiltonian dynamics with
its symplectic geometry. It takes place in the \textit{cotangent bundle} $%
T^{\ast }M_{rob}$, defined as follows. A \textit{dual} notion to the tangent
space $T_{x}M$ to a smooth manifold $M$ at a point $x=(x^i)$ with local is
its {cotangent space} $T_{x}^{\ast }M$ at the same point $x$. Similarly to
the tangent bundle $TM$, for any smooth $n$D manifold $M$, there is
associated its $2n$D \emph{momentum phase-space manifold}, denoted by $%
T^{\ast }M$ and called the \textit{cotangent bundle}. $T^{\ast }M$ is the
disjoint union of all its cotangent spaces $T_{x}^{\ast }M$ at all points $%
x\in M$, i.e., $T^{\ast }M=\dbigsqcup\limits_{x\in M}T_{x}^{\ast }M$.
Therefore, the cotangent bundle of an $n-$manifold $M$ is the vector bundle $%
T^{\ast }M=(TM)^{\ast }$, the (real) dual of the tangent bundle $TM$.
Momentum 1--forms (or, covector-fields) $p_i$ are cross-sections of the
cotangent bundle. Biomechanical \emph{Hamiltonian} (that is, kinetic plus
potential energy) is a natural energy function on the cotangent bundle. The
cotangent bundle $T^*M$ is itself a smooth manifold. It has its own tangent
bundle, $TT^*M$. Cross-sections of the mixed-second bundle $TT^*M$ are the
force 1--forms $F_i= \dot{p}_i$.

There is a unique smooth map from the right-hand branch to the left-hand
branch of the diagram (\ref{covfun}):
\[
TM\ni(x^i,v^i)\mapsto (x^i,p^i)\in T^{\ast }M.
\]
It is called the \emph{Legendre transformation}, or \emph{fiber derivative}
(for details see, e.g. \cite{GaneshSprBig,GaneshADG}).

The fundamental covariant force functor $\mathcal{F}_*:TT^*M \to TTM$ states
that the force 1--form $F_i= \dot{p}_i$, defined on the mixed
tangent--cotangent bundle $TT^*M$, causes the acceleration vector-field $%
a^i= \dot{\bar{v}}^i$, defined on the second tangent bundle $TTM$ of the
configuration manifold $M$. The corresponding \textit{contravariant
acceleration functor} is defined as its inverse map, $\mathcal{F}^*:TTM\to
TT^*M$.

Representation of human motion is rigorously defined in terms of {Euclidean}
$SE(3)$--groups\footnote{%
Briefly, the Euclidean SE(3)--group is defined as a semidirect
(noncommutative) product (denoted by $\rhd$) of 3D rotations and 3D
translations: ~$SE(3):=SO(3)\rhd \Bbb{R}^{3}$. Its most important subgroups
are the following:\newline
{{\frame{$
\begin{array}{cc}
\mathbf{Subgroup} & \mathbf{Definition} \\ \hline
\begin{array}{c}
SO(3),\text{ group of rotations} \\
\text{in 3D (a spherical joint)}
\end{array}
&
\begin{array}{c}
\text{Set of all proper orthogonal } \\
3\times 3-\text{rotational matrices}
\end{array}
\\ \hline
\begin{array}{c}
SE(2),\text{ special Euclidean group} \\
\text{in 2D (all planar motions)}
\end{array}
&
\begin{array}{c}
\text{Set of all }3\times 3-\text{matrices:} \\
\left[
\begin{array}{ccc}
\cos \theta & \sin \theta & r_{x} \\
-\sin \theta & \cos \theta & r_{y} \\
0 & 0 & 1
\end{array}
\right]
\end{array}
\\ \hline
\begin{array}{c}
SO(2),\text{ group of rotations in 2D} \\
\text{subgroup of }SE(2)\text{--group} \\
\text{(a revolute joint)}
\end{array}
&
\begin{array}{c}
\text{Set of all proper orthogonal } \\
2\times 2-\text{rotational matrices} \\
\text{ included in }SE(2)-\text{group}
\end{array}
\\ \hline
\begin{array}{c}
\Bbb{R}^{3},\text{ group of translations in 3D} \\
\text{(all spatial displacements)}
\end{array}
& \text{Euclidean 3D vector space}
\end{array}
$}}}} of full rigid--body motion in all main human joints \cite{TijIJHR}.
The configuration manifold $M$ for human musculo-skeletal dynamics is
defined as a Cartesian product of all included constrained $SE(3)$ groups, $%
M=\prod_{j}SE(3)^{j}$ where $j$ labels the active joints. The configuration
manifold $M$ is coordinated by local joint coordinates $x^i(t),~i=1,...,n=$
total number of active DOF. The corresponding joint velocities $\dot{x}^i(t)$
live in the \emph{velocity phase space} $TM$, which is the \emph{tangent
bundle} of the configuration manifold $M$.

The velocity phase-space $TM$ has the Riemannian geometry with the \textit{%
local metric form}:
\begin{equation}
\langle g\rangle \equiv ds^{2}=g_{ij}dx^{i}dx^{j},  \label{ddg}
\end{equation}
where $g_{ij}=g_{ij}(m,x)$ is the material metric tensor defined by the
biomechanical system's \emph{mass-inertia matrix} and $dx^{i}$ are
differentials of the local joint coordinates $x^{i}$ on $M$. Besides giving
the local distances between the points on the manifold $M,$ the Riemannian
metric form $\langle g\rangle $ defines the system's kinetic energy:
\[
T=\frac{1}{2}g_{ij}\dot{x}^{i}\dot{x}^{j},
\]
giving the \emph{Lagrangian equations} of the conservative skeleton motion
with kinetic-minus-potential energy Lagrangian $L=T-V$, with the
corresponding \emph{geodesic form}
\begin{equation}
\frac{d}{dt}L_{\dot{x}^{i}}-L_{x^{i}}=0,\qquad \text{or}\qquad \ddot{x}%
^{i}+\Gamma _{jk}^{i}\dot{x}^{j}\dot{x}^{k}=0,  \label{geodes}
\end{equation}
where subscripts denote partial derivatives, while $\Gamma _{jk}^{i}$ are
the Christoffel symbols of the affine Levi-Civita connection of the
biomechanical manifold $M$.

The corresponding momentum phase-space $P=T^*M$ provides a natural \emph{%
symplectic structure} that can be defined as follows. As the biomechanical
configuration space $M$ is a smooth $n-$manifold, we can pick local
coordinates $\{dx^{1},...,dx^{n}\}\in M$. Then $\{dx^{1},...,dx^{n}\}$
defines a basis of the cotangent space $T_{x}^*M$, and by writing $\theta
\in T_{x}^*M$ as $\theta =p_{i}dx^{i}$, we get local coordinates $%
\{x^{1},...,x^{n},p_{1},...,p_{n}\}$ on $T^*M$. We can now define the
canonical symplectic form $\omega $ on $P=T^*M$ as:
\[
\omega =dp_{i}\wedge dx^{i},
\]
where `$\wedge $' denotes the wedge or exterior product of exterior
differential forms.\footnote{%
Recall that an \emph{exterior differential form} $\alpha$ of order $p$ (or,
a $p-$\emph{form} $\alpha$) on a base manifold $X$ is a section of the
exterior product bundle $\mathop{\fam0 \w}\limits^p T^*X\to X$. It has the
following expression in local coordinates on $X$
\[
\alpha =\alpha_{\lambda_1\dots\lambda_p} dx^{\lambda_1}\wedge\cdots\wedge
dx^{\lambda_p} \qquad (\mathrm{such~that~~}|\alpha|=p),
\]
where summation is performed over all ordered collections $%
(\lambda_1,...,\lambda_p)$. $\mathbf{\Omega}^p(X)$ is the vector space of $%
p- $forms on a biomechanical manifold $X$. In particular, the 1--forms are
called the \textit{Pfaffian forms}.} This $2-$form $\omega $ is obviously
independent of the choice of coordinates $\{x^{1},...,x^{n}\}$ and
independent of the base point $\{x^{1},...,x^{n},p_{1},...,p_{n}\}\in
T_{x}^*M$. Therefore, it is locally constant, and so $d\omega =0$.\footnote{%
The canonical $1-$form $\theta $ on $T^*M$ is the unique $1-$form with the
property that, for any $1-$form $\beta $ which is a section of $T^*M$ we
have $\beta ^*\theta =\theta $.
\par
Let $f:M\rightarrow M$ be a diffeomorphism. Then $T^*f$ preserves the
canonical $1-$form $\theta $ on $T^*M$, i.e., $(T^*f)^*\theta =\theta $.
Thus $T^*f$ is symplectic diffeomorphism.}

If $(P,\omega )$ is a $2n$D symplectic manifold then about each point $x\in
P $ there are local coordinates $\{x^{1},...,x^{n},p_{1},...,p_{n}\}$ such
that $\omega =dp_{i}\wedge dx^{i}$. These coordinates are called canonical
or symplectic. By the Darboux theorem, $\omega $ is constant in this local
chart, i.e., $d\omega =0$.

\section{Autonomous Hamiltonian Biomechanics}

We develop autonomous Hamiltonian biomechanics on the configuration
biomechanical manifold $M$ in three steps, following the standard symplectic
geometry prescription (see \cite
{GaneshSprSml,GaneshSprBig,GaneshADG,Complexity}):\newline

\noindent \textbf{Step A} Find a symplectic \emph{momentum phase--space} $%
(P,\omega )$.

Recall that a symplectic structure on a smooth manifold $M$ is a
nondegenerate closed\footnote{%
A $p-$form $\beta $ on a smooth manifold $M$ is called \textit{closed} if
its exterior derivative $d=\partial_i dx^i$ is equal to zero,
\[
d\beta=0.
\]
From this condition one can see that the closed form (the \textit{kernel} of
the exterior derivative operator $d$) is conserved quantity. Therefore,
closed $p-$forms possess certain invariant properties, physically
corresponding to the \emph{conservation laws}.
\par
Also, a $p-$form $\beta$ that is an exterior derivative of some $(p-1)-$form
$\alpha$,
\[
\beta=d\alpha,
\]
is called \textit{exact} (the \textit{image} of the exterior derivative
operator $d$). By \emph{Poincar\'e lemma,} exact forms prove to be closed
automatically,
\[
d\beta=d(d\alpha)=0.
\]
\par
Since $d^{2}=0$, \emph{every exact form is closed.} The converse is only
partially true, by Poincar\'{e} lemma: every closed form is \textit{locally
exact}.
\par
Technically, this means that given a closed $p-$form $\alpha \in \Omega
^{p}(U)$, defined on an open set $U$ of a smooth manifold $M$ any point $%
m\in U$ has a neighborhood on which there exists a $(p-1)-$form $\beta \in
\Omega ^{p-1}(U)$ such that $d\beta =\alpha |_{U}.$ In particular, there is
a Poincar\'{e} lemma for contractible manifolds: Any closed form on a
smoothly contractible manifold is exact.} $2-$form $\omega $ on $M$, i.e.,
for each $x\in M$, $\omega (x)$ is nondegenerate, and $d\omega =0$.

Let $T_{x}^*M$ be a cotangent space to $M$ at $m$. The cotangent bundle $%
T^*M $ represents a union $\cup _{m\in M}T_{x}^*M$, together with the
standard topology on $T^*M$ and a natural smooth manifold structure, the
dimension of which is twice the dimension of $M$. A $1-$form $\theta $ on $M$
represents a section $\theta :M\rightarrow T^*M$ of the cotangent bundle $%
T^*M$.

$P=T^*M$ is our momentum phase--space. On $P$ there is a nondegenerate
symplectic $2-$form $\omega $ is defined in local joint coordinates $%
x^{i},p_{i}\in U$, $U$ open in $P$, as $\omega =dx^{i}\wedge dp_{i}$. In
that case the coordinates $x^{i},p_{i}\in U$ are called canonical. In a
usual procedure the canonical $1-$form $\theta $ is first defined as $\theta
=p_{i}dx^{i}$, and then the canonical 2--form $\omega $ is defined as $%
\omega =-d\theta $.

A \emph{symplectic phase--space manifold} is a pair $(P,\omega )$.\newline

\noindent \textbf{Step B} Find a \emph{Hamiltonian vector-field} $X_H$ on $%
(P,\omega)$.

Let $(P,\omega)$ be a symplectic manifold. A vector-field $X:P\rightarrow TP$
is called \emph{Hamiltonian} if there is a smooth function $F:P\to\Bbb{R}$
such that $i_X\omega\,=\,dF$ ($i_X\omega$ denotes the \emph{interior product}
or \emph{contraction} of the vector-field $X$ and the 2--form $\omega$). $X$
is \emph{locally Hamiltonian} if $i_X\omega$ is closed.

Let the smooth real--valued \emph{Hamiltonian function} $H:P\rightarrow \Bbb{%
R}$, representing the total biomechanical energy $H(x,p)\,=\,T(p)\,+\,V(x)$ (%
$T$ and $V$ denote kinetic and potential energy of the system,
respectively), be given in local canonical coordinates $x^{i},p_{i}\in U$, $%
U $ open in $P$. The \emph{Hamiltonian vector-field} $X_{H}$, condition by $%
i_{X_{H}}\omega \,=\,dH$, is actually defined via symplectic matrix $J$, in
a local chart $U$, as
\begin{equation}
X_{H}=J\nabla H=\left( \partial _{p_{i}}H,-\partial _{x^{i}}H\right) ,\qquad
J={\small \left(
\begin{matrix}
0 & I \\
-I & 0
\end{matrix}
\right)} ,  \label{HamVec}
\end{equation}
where $I$ denotes the $n\times n$ identity matrix and $\nabla $ is the
gradient operator.\newline

\noindent \textbf{Step C} Find a \emph{Hamiltonian phase--flow} $\phi _{t}$
of $X_{H}$.

Let $(P,\omega )$ be a symplectic phase--space manifold and $%
X_{H}\,=\,J\nabla H$ a Hamiltonian vector-field corresponding to a smooth
real--valued Hamiltonian function $H:P\rightarrow \Bbb{R}$, on it. If a
unique one--parameter group of diffeomorphisms $\phi _{t}:P\rightarrow P$
exists so that $\frac{d}{dt}|_{t=0}\,\phi _{t}x=J\nabla H(x)$, it is called
the \emph{Hamiltonian phase--flow}.

A smooth curve $t\mapsto \left( x^{i}(t),\,p_{i}(t)\right) $ on $(P,\omega )$
represents an \emph{integral curve} of the Hamiltonian vector-field $%
X_{H}=J\nabla H$, if in the local canonical coordinates $x^{i},p_{i}\in U$, $%
U$ open in $P$, \emph{Hamiltonian canonical equations} hold:
\begin{equation}
\dot{q}^{i}=\partial_{p_{i}} H,\qquad \dot{p}_{i}=-\partial_{q^{i}} H.
\label{HamEq}
\end{equation}

An integral curve is said to be \emph{maximal} if it is not a restriction of
an integral curve defined on a larger interval of $\Bbb{R}$. It follows from
the standard theorem on the \emph{existence} and \emph{uniqueness} of the
solution of a system of ODEs with smooth r.h.s, that if the manifold $%
(P,\omega )$ is Hausdorff, then for any point $x=(x^{i},p_{i})\in U$, $U$
open in $P$, there exists a maximal integral curve of $X_{H}\,=\,J\nabla H$,
passing for $t=0$, through point $x$. In case $X_{H}$ is complete, i.e., $%
X_{H}$ is $C^{p}$ and $(P,\omega )$ is compact, the maximal integral curve
of $X_{H}$ is the Hamiltonian phase--flow $\phi _{t}:U\rightarrow U$.

The phase--flow $\phi _{t}$ is \emph{symplectic} if $\omega $ is constant
along $\phi _{t}$, i.e., $\phi _{t}^*\omega =\omega$

($\phi _{t}^*\omega $ denotes the \emph{pull--back}\footnote{%
Given a map $f:X\to X^{\prime}$ between the two manifolds, the \emph{pullback%
} on $X$ of a form $\alpha$ on $X^{\prime}$ by $f$ is denoted by $f^*\alpha$%
. The pullback satisfies the relations
\begin{eqnarray*}
f^*(\alpha\wedge\beta) =f^*\alpha\wedge f^*\beta, \qquad df^*\alpha
=f^*(d\alpha),
\end{eqnarray*}
for any two forms $\alpha,\beta\in\mathbf{\Omega}^p(X)$.} of $\omega $ by $%
\phi _{t}$),

iff $\frak{L}_{X_{H}}\omega \,=0$

($\frak{L}_{X_{H}}\omega $ denotes the \emph{Lie derivative}\footnote{%
The \textit{Lie derivative} $\frak{L}_u\alpha$ of $p-$form $\alpha$ along a
vector-field $u$ is defined by Cartan's `magic' formula (see \cite
{GaneshSprBig,GaneshADG}):
\[
\frak{L}_u\alpha =u\rfloor d\alpha +d(u\rfloor\alpha).
\]
It satisfies the \emph{Leibnitz relation}
\[
\frak{L}_u(\alpha\wedge\beta)= \frak{L}_u\alpha\wedge\beta +\alpha\wedge%
\frak{L}_u\beta.
\]
Here, the \emph{contraction} $\rfloor$ of a vector-field $u =
u^\m\partial_\m $ and a $p-$form $\alpha =\alpha_{\lambda_1\dots\lambda_p}
dx^{\lambda_1}\wedge\cdots\wedge dx^{\lambda_p}$ on a biomechanical manifold
$X$ is given in local coordinates on $X$ by
\[
u\rfloor\alpha = u^\m \alpha_{\mu\lambda_1\ldots\lambda_{p-1}}
dx^{\lambda_1}\wedge\cdots \wedge dx^{\lambda_{p-1}}.
\]
It satisfies the following relation
\[
u\rfloor(\alpha\wedge\beta)= u\rfloor\alpha\wedge\beta
+(-1)^{|\alpha|}\alpha\wedge u\rfloor\beta.
\]
} of $\omega $ upon $X_{H}$).

Symplectic phase--flow $\phi _{t}$ consists of canonical transformations on $%
(P,\omega )$, i.e., diffeomorphisms in canonical coordinates $x^{i},p_{i}\in
U$, $U$ open on all $(P,\omega )$ which leave $\omega $ invariant. In this
case the \emph{Liouville theorem} is valid: $\phi _{t}$ \emph{preserves} the
\emph{phase volume} on $(P,\omega )$. Also, the system's total energy $H$ is
conserved along $\phi _{t}$, i.e., $H\circ \phi _{t}=\phi _{t}$.

Recall that the Riemannian metrics $g\,=\,<,>$ on the configuration manifold
$M$ is a positive--definite quadratic form $g:TM\rightarrow \Bbb{R}$, in
local coordinates $x^{i}\in U$, $U$ open in $M$, given by (\ref{ddg}) above.
Given the metrics $g_{ij}$, the system's Hamiltonian function represents a
momentum $p$--dependent quadratic form $H:T^*M\rightarrow \Bbb{R}$ -- the
system's kinetic energy $H(p)\,=T(p)=\,\frac{1}{2}<p,p>$, in local canonical
coordinates $x^{i},p_{i}\in U_{p}$, $U_{p}$ open in $T^*M$, given by
\begin{equation}
H(p)=\frac{1}{2}g^{ij}(x,m)\,p_{i}p_{j},  \label{dd19}
\end{equation}
where $g^{ij}(x,m)=g_{ij}^{-1}(x,m)$ denotes the \emph{inverse}
(contravariant) material \emph{metric tensor}
\[
g^{ij}(x,m)\,=\sum_{\chi =1}^{n}m_{\chi }\delta _{rs}\frac{\partial x^{i}}{%
\partial x^{r}}\frac{\partial x^{j}}{\partial x^{s}}.
\]

$T^*M$ is an \emph{orientable} manifold, admitting the standard \emph{volume
form}
\[
\Omega _{\omega _{H}}=\,{\frac{{(-1)^{{\frac{{N(N+1)}}{{2}}}}}}{{N!}}}\omega
_{H}^{N}.
\]

For Hamiltonian vector-field, $X_{H}$ on $M$, there is a base integral curve
$\gamma _{0}(t)\,=\,\left( x^{i}(t),\,p_{i}(t)\right) $ iff $\gamma _{0}(t)$
is a \emph{geodesic}, given by the one--form \emph{force equation}
\begin{equation}  \label{ddmom}
\dot{\bar{p}}_{i}\equiv \dot{p}_{i}+\Gamma
_{jk}^{i}\,g^{jl}g^{km}\,p_{l}p_{m}=0,\qquad \text{with \qquad }\dot{x}%
^{k}=g^{ki}p_{i}.
\end{equation}

The l.h.s $\dot{\bar{p}}_{i}$ of the covariant momentum equation (\ref{ddmom}%
)\ represents the {intrinsic} or {Bianchi covariant derivative} of the
momentum with respect to time $t$. Basic relation $\dot{\bar{p}}_{i}\,=\,0$
defines the \textit{parallel transport} on $T^{N}$, the simplest form of
human--motion dynamics. In that case Hamiltonian vector-field $X_{H}$ is
called the \emph{geodesic spray} and its phase--flow is called the \emph{%
geodesic flow}.

For Earthly dynamics in the gravitational \emph{potential} field $%
V:M\rightarrow \Bbb{R}$, the Hamiltonian $H:T^*M\rightarrow \Bbb{R}$ (\ref
{dd19}) extends into potential form
\[
H(p,x)=\frac{1}{2}g^{ij}p_{i}p_{j}+V(x),
\]
with Hamiltonian vector-field $X_{H}\,=\,J\nabla H$ still defined by
canonical equations (\ref{HamEq}).

A general form of a \emph{driven}, non--conservative Hamiltonian equations
reads:
\begin{equation}
\dot{x}^{i}=\partial _{p_{i}}H,\qquad \dot{p}_{i}=F_{i}-\partial _{x^{i}}H,
\label{FHam}
\end{equation}
where $F_{i}=F_{i}(t,x,p)$ represent any kind of joint--driving \emph{%
covariant torques}, including active neuro--muscular--like controls, as
functions of time, angles and momenta, as well as passive dissipative and
elastic joint torques. In the covariant momentum formulation (\ref{ddmom}),
the non--conservative Hamiltonian equations (\ref{FHam}) become
\[
\dot{\bar{p}}_{i}\equiv \dot{p}_{i}+\Gamma
_{jk}^{i}\,g^{jl}g^{km}\,p_{l}p_{m}=F_{i},\qquad \text{with}\qquad \dot{x}%
^{k}=g^{ki}p_{i}.
\]

The general form of autonomous Hamiltonian biomechanics is given by
dissipative, driven Hamiltonian equations on $T^*M$:
\begin{eqnarray}
\dot{x}^{i} &=&\frac{\partial H}{\partial p_{i}}+\frac{\partial R}{\partial
p_{i}},  \label{fh1} \\
\dot{p}_{i} &=&F_{i}-\frac{\partial H}{\partial x^{i}}+\frac{\partial R}{%
\partial x^{i}},  \label{fh2} \\
x^{i}(0) &=&x_{0}^{i},\qquad p_{i}(0)=p_{i}^{0},  \label{fh3}
\end{eqnarray}
including \textit{contravariant equation} (\ref{fh1}) -- the \textit{%
velocity vector-field}, and \textit{covariant equation} (\ref{fh2}) -- the
\textit{force 1--form} (field), together with initial joint angles and
momenta (\ref{fh3}). Here $R=R(x,p)$ denotes the Raileigh nonlinear
(biquadratic) dissipation function, and $F_{i}=F_{i}(t,x,p)$ are covariant
driving torques of \textit{equivalent muscular actuators}, resembling
muscular excitation and contraction dynamics in rotational form. The
velocity vector-field (\ref{fh1}) and the force $1-$form (\ref{fh2})
together define the generalized Hamiltonian vector-field $X_{H}$; the
Hamiltonian energy function $H=H(x,p)$ is its generating function.

As a Lie group, the biomechanical configuration manifold $%
M=\prod_{j}SE(3)^{j}$ is Hausdorff.\footnote{%
That is, for every pair of points $x_{1},x_{2}\in M$, there are disjoint
open subsets (charts) $U_{1},U_{2}\subset M$ such that $x_{1}\in U_{1}$ and $%
x_{2}\in U_{2}$.} Therefore, for $x\,=\,(x^{i},\,p_{i})\in U_{p}$, where $%
U_{p}$ is an open coordinate chart in $T^*M$, there exists a unique
one--parameter group of diffeomorphisms $\phi_t:T^*M\rightarrow T^*M$, that
is the \emph{autonomous Hamiltonian phase--flow:}
\begin{eqnarray}  \label{fh4}
\phi_t &:&T^*M\rightarrow T^*M:(p(0),x(0))\mapsto (p(t),x(t)), \\
(\phi_t &\circ & \phi _s\, =\,\phi_{t+s},\quad \phi_0\,=\,\text{identity}),
\nonumber
\end{eqnarray}
given by (\ref{fh1}--\ref{fh3}) such that
\[
{\frac{{d}}{{dt}}}\left\vert _{t=0}\right. \phi_tx\,=\,J\nabla H(x).
\]

\section{Time--Dependent Hamiltonian Biomechanics}

In this section we develop time-dependent Hamiltonian biomechanics. For
this, we first need to extend our autonomous Hamiltonian machinery, using
the general concepts of bundles, jets and connections.

\subsection{Biomechanical Bundles and Jets}

While standard autonomous Lagrangian biomechanics is developed on the
configuration manifold $X$, the \emph{time--dependent biomechanics}
necessarily includes also the real time axis $\Bbb{R}$, so we have an \emph{%
extended configuration manifold} $\Bbb{R}\times X$. Slightly more generally,
the fundamental geometrical structure is the so-called \emph{configuration
bundle} $\pi:X\rightarrow \Bbb{R}$. Time-dependent biomechanics is thus
formally developed either on the \emph{extended configuration manifold} $%
\Bbb{R}\times X$, or on the configuration bundle $\pi:X\rightarrow \Bbb{R}$,
using the concept of \textit{jets}, which are based on the idea of \textit{%
higher--order tangency}, or higher--order contact, at some designated point
(i.e., certain anatomical joint) on a biomechanical configuration manifold $%
X $.

\begin{figure}[h]
\centerline{\includegraphics[width=11cm]{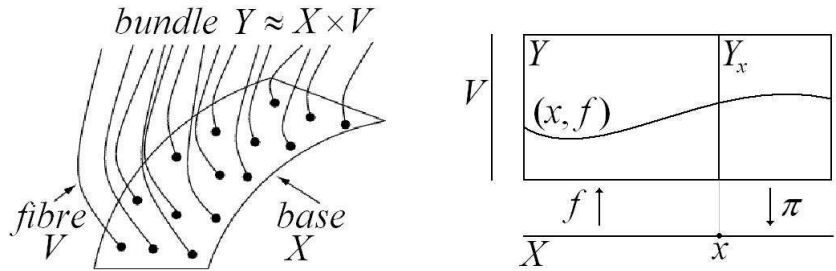}}
\caption{A sketch of a locally trivial fibre bundle $Y\approx X\times V$ as
a generalization of a product space $X\times V$; left -- main components;
right -- a few details (see text for explanation).}
\label{Fibre1}
\end{figure}

In general, tangent and cotangent bundles, $TM$ and $T^{\ast }M$, of a
smooth manifold $M$, are special cases of a more general geometrical object
called \emph{fibre bundle}, denoted $\pi :Y\rightarrow X$, where the word
\emph{fiber} $V$ of a map $\pi :Y\rightarrow X$ is the \emph{preimage} $%
\pi^{-1}(x)$ of an element $x\in X$. It is a space which \emph{locally}
looks like a product of two spaces (similarly as a manifold locally looks
like Euclidean space), but may possess a different \emph{global} structure.
To get a visual intuition behind this fundamental geometrical concept, we
can say that a fibre bundle $Y$ is a \emph{homeomorphic generalization} of a
\emph{product space} $X\times V$ (see Figure \ref{Fibre1}), where $X$ and $V$
are called the \emph{base} and the \emph{fibre}, respectively. $\pi
:Y\rightarrow X$ is called the \emph{projection}, $Y_{x}=\pi ^{-1}(x)$
denotes a fibre over a point $x$ of the base $X$, while the map $f=\pi
^{-1}:X\rightarrow Y$ defines the \emph{cross--section}, producing the
\textit{graph} $(x,f(x))$ in the bundle $Y$ (e.g., in case of a tangent
bundle, $f=\dot{x}$ represents a velocity vector--field).

More generally, a biomechanical configuration bundle, $\pi :Y\to X$, is a
locally trivial fibred (or, projection) manifold over the base $X$. It is
endowed with an atlas of fibred bundle coordinates $(x^\lambda, y^i)$, where
$(x^\la)$ are coordinates of $X$.

\begin{figure}[h]
\centerline{\includegraphics[width=6cm]{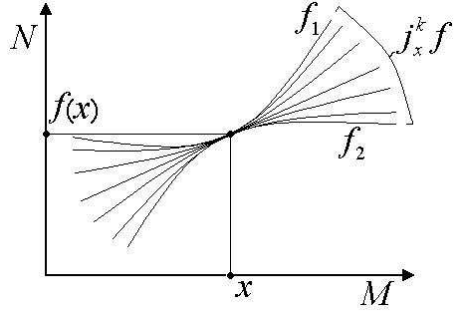}}
\caption{An intuitive geometrical picture behind the $k-$jet concept, based
on the idea of a higher--order tangency (or, higher--order contact). }
\label{jet1}
\end{figure}

Now, a pair of smooth manifold maps, ~$f_{1},f_{2}:M\rightarrow N$~ (see Figure
\ref{jet1}), are said to be $k-$\emph{tangent} (or \emph{tangent of order }$%
k $, or have a $k$th \emph{order contact}) at a point $x$ on a domain
manifold $M$, denoted by $f_{1}\sim f_{2}$, iff
\begin{eqnarray*}
f_{1}(x) &=&f_{2}(x)\qquad \text{called}\quad 0-\text{tangent}, \\
\partial _{x}f_{1}(x) &=&\partial _{x}f_{2}(x),\qquad \text{called}\quad 1-%
\text{tangent}, \\
\partial _{xx}f_{1}(x) &=&\partial _{xx}f_{2}(x),\qquad \text{called}\quad 2-%
\text{tangent}, \\
&&...\qquad \text{etc. to the order }k
\end{eqnarray*}
In this way defined $k-$\emph{tangency} is an \emph{equivalence relation}.

A $k-$\textit{jet} (or, a \emph{jet of order }$k$), denoted by $j_{x}^{k}f$,
of a smooth map $f:M\rightarrow N$ at a point $x\in M$ (see Figure \ref{jet1}%
), is defined as an \emph{equivalence class} of $k-$tangent maps at $x$,
\[
j_{x}^{k}f:M\rightarrow N=\{f^{\prime}:f^{\prime}\text{ is }k-\text{tangent
to }f\text{ at }x\}.
\]

For example, consider a simple function ~$f:X\rightarrow Y,\,x\mapsto y=f(x)$%
, mapping the $X-$axis into the $Y-$axis in $\Bbb{R}^2$. At a chosen point $%
x\in X$ we have:\newline
a $0-$jet is a graph: $(x,f(x))$;\newline
a $1-$jet is a triple: $(x,f(x),f{^{\prime}}(x))$;\newline
a $2-$jet is a quadruple: $(x,f(x),f{^{\prime}}(x),f^{\prime \prime }(x))$,%
\newline
~~ and so on, up to the order $k$ (where $f{^{\prime}}(x)=\frac{df(x)}{dx}$,
etc).\newline
The set of all $k-$jets from $j^k_xf:X\rightarrow Y$ is called the $k-$jet
manifold $J^{k}(X,Y)$.

Formally, given a biomechanical bundle $Y\to X$, its first--order \textit{%
jet manifold} $J^1Y$ comprises the set of equivalence classes $j^1_xs$, $%
x\in X$, of sections $s:X\to Y$ so that sections $s$ and $s^{\prime}$ belong
to the same class iff
\[
Ts\mid _{T_xX} =Ts^{\prime}\mid_{T_xX}.
\]
Intuitively, sections $s,s^{\prime}\in j^1_xs$ are identified by their
values $s^i(x)={s^{\prime}}^i(x)$ and the values of their partial
derivatives $\partial_\mu s^i(x)=\partial_\mu{s^{\prime}}^i(x)$ at the point
$x$ of $X$. There are the natural fibrations \cite{book}
\[
\pi_1:J^1Y\ni j^1_xs\mapsto x\in X, \qquad \pi_{01}:J^1Y\ni j^1_xs\mapsto
s(x)\in Y.
\]
Given bundle coordinates $(x^\la,y^i)$ of $Y$, the associated jet manifold $%
J^1Y$ is endowed with the adapted coordinates
\begin{eqnarray*}
(x^\la,y^i,y_\la^i), \qquad (y^i,y_\la^i)(j^1_xs)=(s^i(x),\partial_\la
s^i(x)), \qquad y^{\prime i}_\la = \frac{\partial x^\m}{\partial{x^{\prime}%
}^\la}(\partial_\m +y^j_\m\partial_j)y^{\prime i}.
\end{eqnarray*}

In particular, given the biomechanical configuration bundle $M\rightarrow
\Bbb{R}$ over the time axis $\Bbb{R}$, the \textit{$1-$jet space} $J^{1}(%
\Bbb{R},M)$ is the set of equivalence classes $j_{t}^{1}s$ of sections $%
s^{i}:\Bbb{R}\rightarrow M$ of the configuration bundle $M\rightarrow \Bbb{R}
$, which are identified by their values $s^{i}(t)$, as well as by the values
of their partial derivatives $\partial _{t}s^{i}=\partial _{t}s^{i}(t)$ at
time points $t\in \Bbb{R}$. The 1--jet manifold $J^{1}(\Bbb{R},M)$ is
coordinated by $(t,x^{i},\dot{x}^{i})$, that is by \textsl{(time,
coordinates and velocities)} at every active human joint, so the 1--jets are
local joint coordinate maps
\[
j_{t}^{1}s:\Bbb{R}\rightarrow M,\qquad t\mapsto (t,x^{i},\dot{x}^{i}).
\]

The \textit{second--order jet manifold} $J^2Y$ of a bundle $Y\to X$ is the
subbundle of $\widehat J^2Y\to J^1Y$ defined by the coordinate conditions $%
y^i_{\lambda\mu}=y^i_{\mu\lambda}$. It has the local coordinates $(x^\la
,y^i, y^i_\la,y^i_{\lambda\leq\mu})$ together with the transition functions
\cite{book}
\begin{eqnarray*}
{y^{\prime}}_{\lambda\mu}^i= \frac{\partial x^\al}{\partial{x^{\prime}}^\m}%
(\partial_\al +y^j_\al\partial_j +y^j_{\nu\alpha}\partial^\nu_j){y^{\prime}}%
^i_\la.
\end{eqnarray*}
The second--order jet manifold $J^2Y$ of $Y$ comprises the equivalence
classes $j_x^2s$ of sections $s$ of\newline
$Y\to X$ such that
\begin{eqnarray*}
y^i_\la (j_x^2s)=\partial_\la s^i(x),\qquad
y^i_{\lambda\mu}(j_x^2s)=\partial_\m\partial_\la s^i(x).
\end{eqnarray*}
In other words, two sections $s,s^{\prime}\in j^2_xs$ are identified by
their values and the values of their first and second--order derivatives at
the point $x\in X$.

In particular, given the biomechanical configuration bundle $M\rightarrow
\Bbb{R}$ over the time axis $\Bbb{R}$, the \textit{$2-$jet space} $J^{2}(%
\Bbb{R},M)$ is the set of equivalence classes $j_{t}^{2}s$ of sections $%
s^{i}:\Bbb{R}\rightarrow M$\ of the configuration bundle $\pi:M\rightarrow
\Bbb{R}$, which are identified by their values $s^{i}(t)$, as well as the
values of their first and second partial derivatives, $\partial
_{t}s^{i}=\partial _{t}s^{i}(t)$ and $\partial _{tt}s^{i}=\partial
_{tt}s^{i}(t)$, respectively, at time points $t\in \Bbb{R}$. The 2--jet
manifold $J^{2}(\Bbb{R},M)$ is coordinated by $(t,x^{i},\dot{x}^{i},\ddot{x}%
^{i})$, that is by \textsl{(time, coordinates, velocities and accelerations)}
at every active human joint, so the 2--jets are local joint coordinate maps%
\footnote{%
For more technical details on jet spaces with their physical applications,
see \cite{book,sard98}).}
\[
j_{t}^{2}s:\Bbb{R}\rightarrow M,\qquad t\mapsto (t,x^{i},\dot{x}^{i},\ddot{x}%
^{i}).
\]

\subsection{Nonautonomous Dissipative Hamiltonian Dynamics}

We can now formulate the time-dependent biomechanics in which the
biomechanical phase space is the Legendre manifold $\Pi $, endowed with the
holonomic coordinates $(t,y^{i},p_{i})$ with the transition functions
\[
p_{i}^{\prime }=\frac{\partial y^{j}}{\partial {y^{\prime }}^{i}}p_{j}.
\]
$\Pi $ admits the canonical form ${\Lambda }$ given by
\[
{\Lambda }=dp_{i}\wedge dy^{i}\wedge dt\otimes \partial _{t}.
\]
We say that a connection
\[
\gamma =dt\otimes (\partial _{t}+\gamma ^{i}\partial _{i}+\gamma
_{i}\partial ^{i})
\]
on the bundle $\Pi \rightarrow X$ is \emph{locally Hamiltonian} if the exterior
form $\gamma \rfloor {\Lambda }$ is closed and Hamiltonian if the form $%
\gamma \rfloor {\Lambda }$ is exact \cite{book}. A connection $\gamma $ is locally
Hamiltonian iff it obeys the conditions:
\[
\partial ^{i}\gamma ^{j}-\partial ^{j}\gamma ^{i}=0,\quad \partial
_{i}\gamma _{j}-\partial _{j}\gamma _{i}=0,\quad \partial _{j}\gamma
^{i}+\partial ^{i}\gamma _{j}=0.
\]

Note that every connection $\Gamma=dt\otimes(\partial_t +\Gamma^i\partial_i)$
on the bundle $Y\to X$ gives rise to the Hamiltonian connection $%
\widetilde\Gamma$ on $\Pi\to X$, given by
\[
\widetilde\Gamma =dt\otimes(\partial_t +\Gamma^i\partial_i
-\partial_j\Gamma^i p_i\partial^j).
\]
The corresponding Hamiltonian form $H_\G$ is given by
\[
H_\G=p_idy^i -p_i\Gamma^idt.
\]

Let $H$ be a \emph{dissipative Hamiltonian form} on $\Pi$, which reads:
\begin{equation}
H=p_idy^i-\mathcal{H} dt=p_idy^i -p_i\Gamma^idt -\widetilde{\mathcal{H}}_\G
dt.  \label{m46}
\end{equation}
We call ${\cal H}$ and $\widetilde{\cal H}$ in the decomposition (\ref
{m46}) the \textit{Hamiltonian} and the \textit{Hamiltonian function}
respectively. Let $\gamma$ be a Hamiltonian connection on $\Pi\to X$
associated with the Hamiltonian form (\ref{m46}). It satisfies the relations
\cite{book,sard98}
\begin{eqnarray}
&&\gamma\rfloor{\Lambda} =dp_i\wedge dy^i+ \gamma_idy^i\wedge dt
-\gamma^idp_i\wedge dt = dH,  \nonumber \\
&&\gamma^i =\partial^i\mathcal{H}, \qquad \gamma_i=-\partial_i\mathcal{H}.
\label{m40}
\end{eqnarray}
From equations (\ref{m40}) we see that, in the case of biomechanics, one and
only one Hamiltonian connection is associated with a given Hamiltonian form.

Every connection $\gamma$ on $\Pi\to X$ yields the system of first--order
differential equations:
\begin{equation}
\dot{y}^i =\gamma^i, \qquad \dot{p}_i =\gamma_i.  \label{m170}
\end{equation}
They are called the \textit{evolution equations}. If $\gamma$ is a
Hamiltonian connection associated with the Hamiltonian form $H$ (\ref{m46}),
the evolution equations (\ref{m170}) become the \emph{dissipative
time-dependent Hamiltonian equations}:
\begin{eqnarray}
\dot{y}^i =\partial^i\mathcal{H}, \qquad \dot{p}_i =-\partial_i\mathcal{H}.
\label{m41}
\end{eqnarray}

In addition, given any scalar function $f$ on $\Pi$, we have the \textit{%
dissipative Hamiltonian evolution equation}
\begin{equation}
d_{H}f=(\partial_t +\partial^i\mathcal{H}\partial_i -\partial_i\mathcal{H}%
\partial^i)\,f,  \label{m59}
\end{equation}
relative to the Hamiltonian $\mathcal{H}$. On solutions $s$ of the
Hamiltonian equations (\ref{m41}), the evolution equation (\ref{m59}) is
equal to the total time derivative of the function $f$:
\begin{eqnarray*}
s^*d_{H}f=\frac{d}{dt}(f\circ s).
\end{eqnarray*}

\subsection{Time--Dependent Biomechanics}

The dissipative Hamiltonian system (\ref{m41})--(\ref{m59}) is the basis for
our time\,\&\,fitness-dependent biomechanics. The scalar function $f$ in (%
\ref{m59}) on the biomechanical Legendre phase-space manifold $\Pi$ is now
interpreted as an \emph{individual neuro-muscular fitness function}. This
fitness function is a `determinant' for the performance of muscular drives
for the driven, dissipative Hamiltonian biomechanics. These muscular drives,
for all active DOF, are given by time\,\&\,fitness-dependent Pfaffian form: $%
F_i=F_i(t,y,p,f)$. In this way, we obtain our final model for
time\,\&\,fitness-dependent Hamiltonian biomechanics: {\large
\begin{eqnarray*}
\dot{y}^i &=& \partial^i\mathcal{H}, \\
\dot{p}_i &=& F_i-\partial_i\mathcal{H}, \\
d_{H}f &= &(\partial_t +\partial^i\mathcal{H}\partial_i -\partial_i\mathcal{H%
}\partial^i)\,f.
\end{eqnarray*}
}

Physiologically, the active muscular drives $F_i=F_i(t,y,p,f)$ consist of
\cite{GaneshSprSml,GaneshWSc}):\newline

\textbf{1. Synovial joint mechanics}, giving the first stabilizing effect to
the conservative skeleton dynamics, is described by the $(y,\dot{y})$--form
of the \textit{Rayleigh--Van der Pol's dissipation function}
\[
R=\frac{1}{2}\sum_{i=1}^{n}\,(\dot{y}^{i})^{2}\,[\alpha _{i}\,+\,\beta
_{i}(y^{i})^{2}],\quad
\]
where $\alpha _{i}$ and $\beta _{i}$ denote dissipation parameters. Its
partial derivatives give rise to the viscous--damping torques and forces in
the joints
\[
\mathcal{F}_{i}^{joint}=\partial R/\partial \dot{y}^{i},
\]
which are linear in $\dot{y}^{i}$ and quadratic in $y^{i}$.\newline

\textbf{2. Muscular mechanics}, giving the driving torques and forces $%
\mathcal{F}_{i}^{musc}=\mathcal{F}_{i}^{musc}(t,y,\dot{ y})$ with $%
(i=1,\dots ,n)$ for human biomechanics, describes the internal {excitation}
and {contraction} dynamics of \textit{equivalent muscular actuators} \cite
{Hatze}.\newline

(a) The \textit{excitation dynamics} can be described by an impulse {%
force--time} relation
\begin{eqnarray*}
F_{i}^{imp} &=&F_{i}^{0}(1\,-\,e^{-t/\tau _{i}})\text{ \qquad if stimulation
}>0 \\
\quad F_{i}^{imp} &=&F_{i}^{0}e^{-t/\tau _{i}}\qquad \qquad \;\quad\text{if
stimulation }=0,\quad
\end{eqnarray*}
where $F_{i}^{0}$ denote the maximal isometric muscular torques and forces,
while $\tau _{i}$ denote the associated time characteristics of particular
muscular actuators. This relation represents a solution of the Wilkie's
muscular \textit{active--state element} equation \cite{Wilkie}
\[
\dot{\mu}\,+\,\Gamma \,\mu \,=\,\Gamma \,S\,A,\quad \mu (0)\,=\,0,\quad
0<S<1,
\]
where $\mu =\mu (t)$ represents the active state of the muscle, $\Gamma $
denotes the element gain, $A$ corresponds to the maximum tension the element
can develop, and $S=S(r)$ is the `desired' active state as a function of the
motor unit stimulus rate $r$. This is the basis for biomechanical force
controller.\newline

(b) The \textit{contraction dynamics} has classically been described by the
Hill's \textit{hyperbolic force--velocity} relation \cite{Hill}
\[
F_{i}^{Hill}\,=\,\frac{\left( F_{i}^{0}b_{i}\,-\,\delta _{ij}a_{i}\dot{y}%
^{j}\,\right) }{\left( \delta _{ij}\dot{y}^{j}\,+\,b_{i}\right) },\,\quad
\]
where $a_{i}$ and $b_{i}$ denote the {Hill's parameters}, corresponding to
the energy dissipated during the contraction and the phosphagenic energy
conversion rate, respectively, while $\delta _{ij}$ is the Kronecker's $%
\delta-$tensor.

In this way, human biomechanics describes the excitation/contraction
dynamics for the $i$th equivalent muscle--joint actuator, using the simple
impulse--hyperbolic product relation
\[
\mathcal{F}_{i}^{musc}(t,y,\dot{y})=\,F_{i}^{imp}\times F_{i}^{Hill}.\quad
\]

Now, for the purpose of biomedical engineering and rehabilitation, human
biomechanics has developed the so--called \textit{hybrid rotational actuator}%
. It includes, along with muscular and viscous forces, the D.C. motor
drives, as used in robotics
\begin{eqnarray*}
&&\mathcal{F}_{k}^{robo}=i_{k}(t)-J_{k}\ddot{y}_{k}(t)-B_{k}\dot{y}%
_{k}(t),\qquad\text{with} \\
&&l_{k}i_{k}(t)+R_{k}i_{k}(t)+C_{k}\dot{y}_{k}(t)=u_{k}(t),
\end{eqnarray*}
where $k=1,\dots,n$, $i_{k}(t)$ and $u_{k}(t)$ denote currents and voltages
in the rotors of the drives, $R_{k},l_{k}$ and $C_{k}$ are resistances,
inductances and capacitances in the rotors, respectively, while $J_{k}$ and $%
B_{k}$ correspond to inertia moments and viscous dampings of the drives,
respectively.

Finally, to make the model more realistic, we need to add some \textit{%
stochastic torques and forces}:
\[
\mathcal{F}_{i}^{stoch}=B_{ij}[y^{i}(t),t]\,dW^{j}(t),
\]
where $B_{ij}[y(t),t]$ represents continuous stochastic \textit{diffusion
fluctuations}, and $W^{j}(t)$ is an $N-$variable \textit{Wiener process}
(i.e., generalized Brownian motion) \cite{StrAttr}, with
\[
dW^{j}(t)=W^{j}(t+dt)-W^{j}(t),\qquad (\text{for} ~~j=1,\dots,n=\text{no. of
active DOF}).
\]

\section{Conclusion}

In this paper we have proposed the time-dependent Hamiltonian form of
human biomechanics. Starting with the Covariant Force Law: $F_{i}=m_{ij}a^{j}$ on the biomechanical configuration manifold $M$, we have first developed the
autonomous Hamiltonian biomechanics:
$$
\dot{x}^{i} =\frac{\partial H}{\partial p_{i}}+\frac{\partial R}{\partial
p_{i}}, \qquad
\dot{p}_{i} =F_{i}-\frac{\partial H}{\partial x^{i}}+\frac{\partial R}{%
\partial x^{i}},
$$
on the symplectic phase space that is the cotangent bundle $TM$ of $M$.
Then we have introduced powerful
geometrical machinery consisting of fibre bundles and jet manifolds associated to the biomechanical manifold $M$. Using the jet formalism, we derived
time-dependent, dissipative, Hamiltonian equations:
$$
\dot{y}^i = \partial^i\mathcal{H}, \qquad
\dot{p}_i = F_i-\partial_i\mathcal{H},
$$
together with the fitness evolution
equation:
$$
d_{H}f = (\partial_t +\partial^i\mathcal{H}\partial_i -\partial_i\mathcal{H%
}\partial^i)\,f.
$$
for the general time-dependent human biomechanical
system.

\end{document}